\documentclass[a4paper,11pt]{article}
\pdfoutput=1 % if your are submitting a pdflatex (i.e. if you have
\usepackage{jcappub} % for details on the use of the package, please
\usepackage[T1]{fontenc} % if needed

\usepackage{graphicx}
\usepackage{epstopdf}
\usepackage{caption}
\usepackage{graphicx}
\usepackage{bm}
\usepackage[latin1,utf8]{inputenc}
\usepackage{amsmath}
\usepackage{subfigure}
\usepackage{amssymb}
\usepackage{makeidx}
\usepackage{color}
\usepackage{adjustbox}
\usepackage{lipsum}
\usepackage{mathrsfs}
\usepackage{multirow}
\usepackage{amsfonts}

\title{ Dark energy with a gradient coupling to the dark matter fluid: cosmological dynamics and structure formation }

\author[a,b]{Jibitesh~Dutta,}
\author[c,d]{Wompherdeiki Khyllep,}
\author[e]{Nicola Tamanini}

\affiliation[a]{Mathematics Division, Department of Basic Sciences and Social Sciences, North Eastern Hill University,
NEHU Campus, Shillong, Meghalaya 793022, India}
\affiliation[b]{Inter University Centre for Astronomy and Astrophysics, Pune 411 007, India}
\affiliation[c]{Department of Mathematics,~ North Eastern Hill
University,~NEHU Campus, Shillong, Meghalaya 793022, India}
\affiliation[d]{Department of Mathematics, St. Anthony's College, Shillong, Meghalaya 793001, India}
\affiliation[e]{Institut de Physique Th\'eorique, CEA-Saclay,
CNRS UMR 3681, Universit\'e Paris-Saclay, F-91191 Gif-sur-Yvette, France}
\emailAdd{jdutta29@gmail.com, jdutta@associates.iucaa.in}
\emailAdd{sjwomkhyllep@gmail.com}
\emailAdd{nicola.tamanini@cea.fr}

\abstract{
We consider scalar field models of dark energy interacting with dark matter through a coupling proportional to the contraction of the four-derivative of the scalar field with the four-velocity of the dark matter fluid.
The coupling is realized at the Lagrangian level employing the formalism of Scalar-Fluid theories, which use a consistent Lagrangian approach for relativistic fluid to describe dark matter.
This framework produces fully covariant field equations, from which we can derive unequivocal cosmological equations at both background and linear perturbations levels.
The background evolution is analyzed in detail applying dynamical systems techniques, which allow us to find the complete asymptotic behavior of the universe given any set of model parameters and initial conditions.
Furthermore we study linear cosmological perturbations investigating the growth of cosmic structures within the quasi-static approximation.
We find that these interacting dark energy models give rise to interesting phenomenological dynamics, including late-time transitions from dark matter to dark energy domination, matter and accelerated scaling solutions and dynamical crossing of the phantom barrier.
Moreover we obtain possible deviations from standard $\Lambda$CDM behavior at the linear perturbations level, which have an impact on the dynamics of structure formation and might provide characteristic observational signatures.
}

\begin{document}

\maketitle

\section{Introduction}\label{sec:Intro}

Although various astrophysical observations have by now confirmed the present accelerated expansion of our universe \cite{Riess:1998cb,Perlmutter:1998np,Betoule:2014frx,Ade:2013zuv,Ade:2015xua}, a search for the exact nature of the phenomenon driving this acceleration is still under way.
In order to find a theoretical explanation, two main approaches are usually considered: modifying the gravitational part of the Einstein equations \cite{Randall:1999ee,Randall:1999vf,Dvali:2000hr} or introducing the concept of dark energy (DE) as a new mysterious cosmological component.
The time-independent cosmological constant $\Lambda$ is known to be the simplest DE candidate proposed so far, and it appears to be consistent with the current observations.
$\Lambda$ has however its own theoretical problems, specifically the so called cosmological constant problem and the cosmic coincidence problem \cite{Weinberg:1988cp,Martin:2012bt,Steinhardt:1999nw}.
In order to overcome these issues, another promising explanation for DE that has been proposed is a dynamical scalar field with a self-interacting potential which can mimic the cosmological constant behaviour at late times (see \cite{Copeland:2006wr,Tsujikawa:2013fta} for reviews).
Scalar field models can moreover be well motivated by the low energy limit of well known high energy theories, for example string theory. 

Once DE is allowed to be characterised by a dynamical field, it is natural to consider possible interactions with other cosmological components. An interaction between DE and ordinary matter is severely constrained by solar system experimental tests probing the magnitude of a possible fifth force \cite{Will,Will:2014kxa}.
Nevertheless a coupling between DE and dark matter (DM) cannot be excluded by Solar System experiments and cosmological observations still allow such possibility (see e.g.~\cite{Bolotin:2013jpa,Valiviita:2015dfa,Wang:2016lxa}).
The presence of an interaction in the dark sector is an interesting hypothesis in modern cosmology, which could have an effect on both background and perturbation dynamics \cite{Gumjudpai:2005ry,CalderaCabral:2009ja}.
One of the main features of interacting DE is the existence of late time accelerated scaling attractors, which in principle can alleviate the coincidence problem \cite{Wetterich:1994bg,Amendola:1999er,Zimdahl:2001ar}.
Furthermore current observations \cite{Ade:2013zuv,Ade:2015xua} mildly suggest that the effective equation of state (EoS) parameter of DE might be smaller than that of a cosmological constant, i.e.~smaller than $-1$.
This phenomenon cannot be achieved in uncoupled canonical models with a single scalar field.
A possible solution is provided by a phantom scalar field, which however suffers from classical and quantum instabilities \cite{Carroll:2003st,Cline:2003gs}.
Interestingly, the crossing of the phantom divide line (i.e.~the crossing of the value $-1$ for the EoS parameter) can also be achieved through an interaction in the dark sector, without resorting to a phantom scalar field.
Some current observational datasets provide moreover some indications of a non-vanishing late time interaction \cite{Salvatelli:2014zta,Valiviita:2015dfa,vandeBruck:2016hpz}, and future experiments will be able to better constrain such hypothesis \cite{Cai:2015zoa,Caprini:2016qxs,Cai:2017yww}.

A coupling between DE and DM could thus be used to alleviate the cosmic coincidence problem and to explain a possible excursion in the phantom regime.
The problem then arises on how to define this interaction from a theoretical perspective.
Due to the unknown fundamental nature of both DE and DM, any coupling proposed in the literature can only be defined phenomenologically at the level of the field equations (see e.g.~\cite{Boehmer:2008av,Boehmer:2009tk,Dutta:2016dnt}).
This has created some problems whenever extensions of these models from the background dynamics to the perturbation level or full covariant level were considered \cite{Valiviita:2008iv,Faraoni:2014vra,Skordis:2015yra}, and it has moreover limited the theoretical framework describing a possible dark interaction \cite{Tamanini:2015iia}.
In order to overcome these issues, a new framework of coupled DE models, generally known as Scalar-Fluid theories, was introduced in \cite{Boehmer:2015kta,Boehmer:2015sha}.
In this class of theories a consistent variational approach is employed by modelling DM as a dynamical fluid using Brown's Lagrangian formulation of relativistic fluids \cite{Brown:1992kc}.
This allows for a well-defined, though still phenomenological, Lagrangian, able to provide fully covariant equations of motion, and thus unequivocal cosmological dynamics at both background and perturbation levels.
Scalar-Fluid theories include and extend most of the previously considered interactions between scalar field DE and DM.
Within their framework the scalar field's and the fluid's degrees of freedom can not only be coupled algebraically \cite{Boehmer:2015kta}, but interacting terms between the scalar field gradient (its derivative) and the fluid's four-velocity can be created at the Lagrangian level \cite{Boehmer:2015sha}.
For some applications of Scalar-Fluid theories we refer the reader to \cite{Koivisto:2015qua,Boehmer:2015ina,Brax:2015fcf,Tamanini:2016klr,Dutta:2017kch}.

In the present work, we investigate the cosmological dynamics of Scalar-Fluid DE models with the gradient (derivative) coupling introduced in \cite{Boehmer:2015sha}.
We consider an arbitrary self-interacting scalar field potential and make use of dynamical system methods to characterise the background cosmic evolution in detail.
Dynamical system techniques constitute a powerful tool to determine the asymptotic behavior of any cosmological model.
The objective is to relate the critical points of the phase space with important cosmological periods, for example inflation, matter dominated and accelerated DE dominated eras.
A similar analysis has already been performed in \cite{Boehmer:2015sha}, where however only exponential potentials were considered for the scalar field.
Apart from a mathematical point of view, the generalization to arbitrary scalar field potentials, is also well motivated by the low-energy limit of more fundamental high-energy theories, as well as by comparison with different phenomenological models of DE \cite{Copeland:2006wr,Tsujikawa:2013fta}. In order to analyze the cosmological dynamics for arbitrary potentials, we rely on the method developed in \cite{Fang:2008fw} for the quintessence field, and subsequently applied in the context of $k$-essence \cite{Dutta:2016bbs}, braneworld theories \cite{Leyva:2009zz,Escobar:2011cz,Escobar:2012cq,Dutta:2016dnt}, tachyon fields \cite{Quiros:2009mz,Fang:2010zze,Farajollahi:2011ym}, quintom fields \cite{Leon:2012vt} and loop quantum gravity \cite{Xiao:2011nh}. We also discuss the stability of non-hyperbolic critical points (critical points whose Jacobian matrix present eigenvalues of vanishing real part), for which standard linear stability theory fails to determine their properties.
Non-hyperbolic critical points arise in the background dynamics of different DE models and might contain important information and interesting dynamical features regarding the late time universe.
To determine their stability, we either use some advanced mathematical tool such as center manifold theory \cite{Wiggins,Perko,Boehmer:2011tp,Fang:2008fw,TamaniniPhDthesis} or numerical computational techniques such as the analysis of perturbed trajectories near the critical point \cite{Dutta:2016dnt,Dutta:2016bbs}.

Furthermore, in order to check the viability of these coupling models during the formation of large scale structures, we investigate them at the linear cosmological perturbation level.
Cosmological perturbations for Scalar-Fluid theories have been first investigated in \cite{Koivisto:2015qua,Boehmer:2015ina}.
In the quasi-static approximation, we used to describe the structure dynamics deep inside the horizon, new interesting modifications to the equations describing the growth of structures appear, with properties which cannot easily be obtained in other models of DE or even modified gravity.
In the present paper we study the dynamics of scalar perturbations during eras of effective matter domination, where the scalar field might imprint particular features during the growth of structure, while not affecting the evolution at the background level.
This analysis can be useful to identify specific signatures of Scalar-Fluid theories to look for by cosmological observations.

The organization of this paper is as follows.
In Sec.~\ref{sec:scalar}, we briefly review the theoretical framework of Scalar-Fluid theories introduced in \cite{Boehmer:2015sha}, presenting the basic cosmological equations.
In Sec.~\ref{sec:background} we explore the dynamics at the background cosmological level, starting by deriving an autonomous system of differential equations in a spatially flat homogeneous and isotropic universe.
In Secs.~\ref{sec:model1} and \ref{sec:model2}, we then consider two specific models, corresponding to two different coupling functions depending on the scalar field gradient and the fluid's four velocity, and explore their dynamics using dynamical system techniques.
In Sec.~\ref{sec:perturbation}, we investigate the implications of these derivative coupling models in the growth of cosmological structures using linear perturbation theory within the quasi-static approximation.
Finally, we draw our conclusions in Sec.~\ref{sec:conc}.

\par {\it Notation}: In this work, we assume the $(-,+,+,+)$ signature convention for the metric. We shall adopt units where $8\pi G=c=\hslash=1$. Moreover, the comma notation denotes standard partial derivatives (i.e. $\phi_{,\mu}=\partial_{\mu}\phi$).

\section{Scalar-Fluid theories with a derivative coupling}
\label{sec:scalar}

\subsection{The Scalar-Fluid action}

The total action of scalar-fluid theories is given by \cite{Boehmer:2015kta,Boehmer:2015sha}
\begin{equation}\label{action}
S=\int d^4x\left[\mathcal{L}_{\rm grav}+\mathcal{L}_{\rm mat}+\mathcal{L}_{\phi}+\mathcal{L}_{\rm int}\right],
\end{equation}
where $\mathcal{L}_{\rm grav}$ stands for the gravitational Lagrangian, $\mathcal{L}_{\rm mat}$ stands for the matter Lagrangian, $\mathcal{L}_{\phi}$ stands for the scalar field Lagrangian and $\mathcal{L}_{\rm int}$ stands for the interacting Lagrangian.
The gravitational Lagrangian $\mathcal{L}_{\rm grav}$ is given by the standard Einstein-Hilbert Lagrangian
\begin{equation}\label{Lgrav}
\mathcal{L}_{\rm grav}=\frac{\sqrt{-g}}{2}R,
\end{equation}
where $g$ is the determinant of the metric $g_{\mu\nu}$ and $R$ is the Ricci scalar.
The matter Lagrangian $\mathcal{L}_{\rm mat}$ for relativistic fluid  \cite{Brown:1992kc} is given by
\begin{equation}\label{Lmat}
\mathcal{L}_{\rm mat}=-\sqrt{-g} \rho(\mathfrak{n},\mathfrak{s})+J^{\mu}\left(\varphi_{,\mu}+\mathfrak{s}\, \theta_{,\mu}+\beta_A \alpha^A_{,\mu}\right),
\end{equation}
where $\rho(\mathfrak{n},\mathfrak{s})$ is the energy density of the fluid, considered to be depending only on the particle number density $\mathfrak{n}$ and the entropy density per particle $\mathfrak{s}$.
Here $\theta,\,\varphi$ and $\beta_A$ are Lagrange multipliers with $A=1,\,2,\,3$, and $\alpha_A$ are the Lagrangian coordinates of the fluid.
The quantity $J^{\mu}$, which denotes the vector density particle number, is connected to $\mathfrak{n}$ as
\begin{equation}\label{Jmun}
J^{\mu}=\sqrt{-g}\,\mathfrak{n}\,u^{\mu},\quad |J|=\sqrt{-g_{\mu\nu} J^{\mu}J^{\nu}},\quad \mathfrak{n}=\frac{|J|}{\sqrt{-g}},
\end{equation}
where $u^{\mu}$ is the fluid 4-velocity satisfying $u_{\mu}u^{\mu}=-1$.
The scalar field Lagrangian $\mathcal{L}_{\phi}$ is given by the canonical form
\begin{equation}\label{Lphi}
\mathcal{L}_{\phi}=-\sqrt{-g}\left[\frac{1}{2}\partial_{\mu}{\phi}\partial^{\mu}{\phi}+V(\phi)\right],
\end{equation}
where $V$ stands for an arbitrary potential of the scalar field $\phi$.
Finally we consider the interacting Lagrangian term $\mathcal{L}_{\rm int}$ where the scalar field's first spacetime derivative $\partial_{\mu}\phi$ interacts with the fluid's degrees of freedom as \cite{Boehmer:2015sha}
\begin{equation}\label{der_Lag}
\mathcal{L}_{\rm int}=f(\mathfrak{n},\mathfrak{s},\phi) J^\mu \partial_\mu \phi,
\end{equation}
where $f(\mathfrak{n},\mathfrak{s},\phi)$ is an arbitrary function.
Note that this term represents an effective coupling between the gradient of the scalar field $\partial_{\mu}\phi$ and the fluid's 4-velocity $u^{\mu}$ as
\begin{equation}
	\mathcal{L}_{\rm int}= \sqrt{-g}\, \mathfrak{n}\, f(\mathfrak{n},\mathfrak{s},\phi)\, u^\mu \partial_\mu \phi \,.
\end{equation}

\subsection{Background cosmological equations}

In what follows we shall first consider the cosmological evolution of the universe based on the action (\ref{action}) under a spatially flat, homogeneous and isotropic Friedmann-Robertson-Walker (FRW) universe, described by the metric
 \begin{align} \label{metric}
 ds^2=-dt^2+a^2(t)(dx^2+dy^2+dz^2),
\end{align}
where $a(t)$ is the scale factor depending on the cosmic time $t$ and $x$, $y$, $z$ are Cartesian coordinates.
The cosmological equations obtained from action (\ref{action}), are given by
\begin{align}
3H^2 &= \rho+\rho_{\phi}+\rho_{\rm int} \,, \label{Fried_Der} \\
2 \dot{H}+3 H^2 &= -\left(p+p_{\phi}+p_{\rm int}\right) \,, \label{acc_eqn_er} 
\end{align}
where $H=\frac{\dot{a}}{a}$ is the Hubble parameter and an over-dot denotes the time derivative.
In the equations above, $\rho$ and $p$ denote the energy density and pressure of the matter fluid with a linear EoS (EoS) $w$ defined by $p=w \rho$ (we will mainly consider $w=0$ in what follows, i.e.~non-relativistic matter). % ($-1 \leq w \leq 1$).
$\rho_{\rm int}$ and $p_{\rm int}$ are the interacting energy density and pressure, while $\rho_{\phi}$ and $p_{\phi}$ are the scalar field energy density and pressure.
They are respectively given by \cite{Boehmer:2015sha}
\begin{align}
\rho_{\rm int}=0, ~~p_{\rm int}=-\mathfrak{n}^2\frac{\partial f}{\partial \mathfrak{n}}\dot{\phi},~~\rho_{\phi}=\frac{1}{2}\dot{\phi}^2+V,~~p_{\phi}&=\frac{1}{2}\dot{\phi}^2-V \,.
\end{align}
Finally varying action (\ref{action}) with respect to the scalar field $\phi$, we obtain the modified Klein-Gordon equation as
\begin{align}\label{Klein_Der2}
\ddot{\phi}+3H\dot{\phi}+\frac{\partial V}{\partial \phi}-3 H \mathfrak{n}^2\frac{\partial f}{\partial \mathfrak{n}}=0 \,.
\end{align}
Note that the Friedmann equation (\ref{Fried_Der}) does not get modified by the coupling term since $\rho_{\rm int}=0$;
whereas the acceleration equation (\ref{acc_eqn_er}) and scalar field equation (\ref{Klein_Der2}) are affected by the interaction.

\section{Background cosmological dynamics}
\label{sec:background}

\subsection{Formation of the autonomous system}

We employ the following dimensionless variables to convert the cosmological equations (\ref{Fried_Der})-(\ref{Klein_Der2}) to an autonomous system of equations,
\begin{align}
\sigma=\frac{\sqrt{\rho}}{\sqrt{3} H}\,, \quad x=\frac{\dot\phi}{\sqrt{6} H}\,,\quad
y=\frac{\sqrt{V}}{\sqrt{3}H}\,,\quad s=-\frac{1}{V}\frac{d V}{d \phi},
\label{variable1}
\end{align}
where the variable $s$ is usually employed for arbitrary self-interacting potentials \cite{Fang:2008fw,TamaniniPhDthesis}.
Using the dimensionless variables (\ref{variable1}), the Friedmann equation (\ref{Fried_Der}) becomes
\begin{align}
1=\sigma^2+x^2+y^2.
\label{constraint_der}
\end{align}
This acts as a constraint equation for the dimensionless variables (\ref{variable1}), effectively decreasing the dimension of the phase space by one.
The DE density parameter and the DM energy density parameter are respectively given by
  \begin{align}
  \Omega_{\phi} &\equiv \frac{\rho_\phi}{3H^2} = x^2+y^2 \,,\\
  \Omega_m &\equiv \frac{\rho}{3 H^2} = 1-x^2-y^2 \,.
  \end{align}
Using the dimensionless variables (\ref{variable1}), the cosmological equations (\ref{Fried_Der})-(\ref{Klein_Der2}) can be rewritten as the following autonomous system of equations
\begin{align}
 x' &= -\frac{1}{2} \left[ 3 x ((w-1) x^2+(w+1)y^2+1-w)-\sqrt{6}(A(x^2-1)+s y^2) \right], \label{x_der}\\
  y' &= -\frac{1}{2} y \left[ 3 \left((w-1) x^2 +  (w+1) (y^2-1)\right)+\sqrt{6} x  (s-A)\right], \label{y_der}\\
  s'&=-\sqrt{6}\, x\, g(s),\label{s_der}
\end{align}
where $g(s)=s^2(\Gamma(s)-1)$ and
\begin{align}\label{Gamma}
\Gamma=V\frac{d^2V}{d \phi^2}\left(\frac{d V}{d \phi}\right)^{-2},
\end{align}
while a prime denotes differentiation with respect to the number of $e$-folds $N$ defined by $dN=H dt$.
In Eqs. (\ref{x_der})-(\ref{s_der}), we have also introduced the dimensionless quantity
\begin{align}
A=-\frac{1}{H} \mathfrak{n}^2 \frac{\partial f}{\partial \mathfrak{n}} \,.
\label{eq:def_A}
\end{align}
Different types of scalar field potential $V(\phi)$ lead to different forms of $\Gamma$, with the exponential potential being the simplest possible choice ($\Gamma=1$).
In what follows we will assume that $\Gamma$ can always be written as a function of $s$.
This assumption holds for a wide class of scalar field potential, including many of the cases considered in cosmology (see e.g.~\cite{TamaniniPhDthesis}).
In order to close the system, we must also specify the coupling function $f$ from which $A$ can be obtained.
For some choices of $f$, the quantity $A$ depends solely on $x$, $y$ and $s$ and the resulting system remains autonomous.
However in the most general situation $A$ cannot be written as a function of $x$, $y$ and $s$ only, and an extra variable must be introduced, increasing in this way the dimension of the system.
In what follows, we consider the two choices of $f$ given in Table \ref{models}, referring to them as Model~I and Model~II.
They generalise the cases considered in \cite{Boehmer:2015sha}.
In Model~I, the quantity $A$ depends solely on $x$, $y$ and $s$, while in Model~II it cannot be written in terms of $x$, $y$ and $s$ only and a further variable must be introduced (see Sec.~\ref{sec:model2}).

\begin{table}%[!ht]
\centering
\begin{tabular}{|c|c|c|}
\hline
\mbox{} & $f$ & $A$ \\
\hline
& \multirow{3}{*}{$ -\gamma \left(-\frac{1}{V}\frac{dV}{d\phi}\right)^\beta \rho^{1/2-\alpha}V^{\alpha}/\sqrt{3}n$} & \\
Model I & & $\xi\, s^\beta \, y^{2\alpha}(1-x^2-y^2)^{1/2-\alpha}$ \\
& & \\
\hline
& \multirow{3}{*}{$\xi \left(-\frac{1}{V}\frac{dV}{d\phi}\right)^\beta \frac{H_0}{n}$} & \\
Model II &   & $\xi\, s^\beta\, \frac{H_0}{H}$\\
& &\\
\hline
\end{tabular}
\caption{Explicit expression of $A$ (see Eq.~\eqref{eq:def_A}) for the choices of the interacting function $f$ considered in Sec.~\ref{sec:model1} (Model~I) and Sec.~\ref{sec:model2} (Model~II). Here $\alpha$, $\beta$, $\gamma$, $\xi$ are all dimensionless parameters and $\gamma$ is defined as $\gamma=\xi / \left[(\frac{1}{2}-\alpha)(w+1)-1\right]$.}
\label{models}
\end{table}

%%%%%%%%%%%%%%%%%%%%%%%%%%%%%%%%%%%%%%%%%%%%%%%%%%%%%%%%%%%%%%%%%%%%%%%%%%%%%%%%%%%%%%%%%%%%%%%%%%%%%%%%%%%%%%%%%%%%%%%%%%%%%%%%

\subsection{Model I}
\label{sec:model1}

This section deals with the phase space analysis of the dynamical system~(\ref{x_der})-(\ref{s_der}) for Model~I.
In terms of the dimensionless variables (\ref{variable1}), the acceleration equation (\ref{acc_eqn_er}) can be expressed as
\begin{align}
\frac{\dot{H}}{H^2}=\frac{3}{2}\left\lbrace -(1+w)+(w-1)x^2+(w+1)y^2-\sqrt{\frac{2}{3}}\, x\,\xi\, s^\beta \, y^{2\alpha}(1-x^2-y^2)^{1/2-\alpha}\right\rbrace \,.
\end{align}
This implies that the effective EoS parameter $w_{\rm eff}$ can be written as
\begin{align}
  w_{\rm eff} &\equiv\frac{p+\frac{1}{2}\dot{\phi}^2-V+p_{\rm int}}{\rho+\frac{1}{2}\dot{\phi}^2+V+\rho_{\rm int}}\nonumber\\
  & =x^2-y^2+w(1-x^2-y^2)+\sqrt{\frac{2}{3}}\, x\,\xi\, s^\beta \, y^{2\alpha}(1-x^2-y^2)^{1/2-\alpha} \,.
\end{align}
The physically meaningful assumption $\rho\geq 0$, namely $\sigma^2\geq 0$, implies that one obtains from the constraint equation (\ref{constraint_der}) the condition
\begin{align}
x^2+y^2\leq 1 \,.
\end{align}
Hence, the three dimensional phase space of the system (\ref{x_der})-(\ref{s_der}) for Model~I is given by
\begin{equation}
\Psi=\left\lbrace(x,y)\in \mathbb{R}^2:0\leq x^2+y^2\leq 1\right\rbrace \times \left\lbrace s \in \mathbb{R}\right\rbrace.
\end{equation}
Note also that whenever $\alpha$ is an integer number the dynamical system (\ref{x_der})-(\ref{s_der}) is invariant under the transformation $y \mapsto -y$, implying that the dynamics for $y>0$ can be mirrored into the dynamics for $y<0$.

\begin{table}%[!ht]
\centering
\begin{tabular}{|c|c|c|c}
\hline
\multirow{4}{*}{Model I} & \multirow{2}{*}{$\beta=0$} & $\alpha=0$\\
\cline{3-2}\cline{4-2}
& & $\alpha=1/2$ \\
\cline{2-2}\cline{3-2}\cline{4-2}
& \multirow{2}{*}{$\beta=1$} & $\alpha=0$ \\
\cline{3-2}\cline{4-2}
& & $\alpha=1/2$\\
\hline
\multirow{2}{*}{Model II} & $\beta=0$ & -- \\
\cline{2-2}\cline{3-2}\cline{4-2}
 &  $\beta=1$ & -- \\
\hline
\end{tabular}
\caption{Choices of $\alpha$ and $\beta$ considered in this paper for both Model~I and II (see Table~\ref{models}).}
\label{model_I}
\end{table}

Model~I becomes singular whenever $y=0$, $\sigma=0$ or $s=0$ for all values of $\alpha$, $\beta$ except in the range $0\leq \alpha \leq \frac{1}{2}$ and $\beta \geq 0$.
For this reason we will restrict our study only to values of $\alpha$ and $\beta$ belonging to this range.
In general it is difficult to perform a complete dynamical system analysis for arbitrary values of $\alpha$ and $\beta$, so in what follows we will investigate the background cosmological dynamics only for the specific choices of $\alpha$ and $\beta$ reported in Table~\ref{model_I}.
We shall denote a particular model with given $\beta$ and $\alpha$ by $(\beta,\alpha)$.
For example, the model with $\beta=1$, $\alpha=0$ will be denoted as the $(1,0)$ model.
The background cosmological dynamics for the $(0, \frac{1}{2})$ model coincides with the one obtained by the $k$-essence scalar field studied in \cite{Nicola, Dutta:2016bbs}, and for this reason it will not be present again here.

%%%%%%%%%%%%%%%%%%%%%%%%%%%%%%%%%%%%%%%%%%%%%%%%%%%%%%%%%%%%%%%%%%

\subsubsection{$(0,0)$ model}
\label{subsec:00model}

\begin{table}%[!ht]
\centering
\begin{tabular}{cccccc}
  \hline\hline
  Point~ & ~~~~$x$~~~~ & ~~~~$y$ ~~~~&~~~~$s$~~~~&Existence&~~~~$w_{\rm eff}$\\%~~&~~Acceleration\\
  \hline

  $A_{1 \pm}$ & $\pm 1$ & 0 &$s_*$ &Always&1\\[1.5ex]%&No\\

  $A_2$ & $\frac{s_*}{\sqrt{6}}$ & $\sqrt{1-\frac{s_*^2}{6}}$ &$s_*$ &$s_*^2 \leq 6$&$\frac{s_*^2-3}{3}$\\[1.5ex]

  $A_3$ & $-\sqrt{\frac{2}{3(w-1)^2+2\xi^2}} \xi$ &
  0 &$s_*$&Always&$\frac{3w}{3+2\xi^2}$\\[1.5ex]

   $A_4$ & $0$ & $1$ &$0$ &Always&$-1$\\[1.5ex]
  \hline\hline
\end{tabular}
%\end{adjustbox}
\caption{Critical points of $(0,0)$ model (Sec.~\ref{subsec:00model}).}
\label{tab:der_A}
\end{table}

\begin{table}%[!ht]
\centering
Here: $\Delta=\sqrt{\frac{3}{3(1-w)^2+2\xi^2}}$\\
\begin{adjustbox}{width=1\textwidth}
\small
\begin{tabular}{ccccc}

\hline\hline
\mbox{Point} &~~~~~~~~~~~~$\lambda_1$~~~~~~~~~ &~~~~~~~~~~~~ $\lambda_2$~~~~~~~~~~~~~~~~~~~~ &$\lambda_3$ ~~~~~~~~~~~~~~~~~& ~~~~~~~~~Stability \\
\hline  \\[0.1ex]

$A_{1 \pm}$&$3(1-w)$&$3\mp \frac{\sqrt{6}s_*}{2}$&$\mp \sqrt{6}\,dg(s_*)$&Unstable node/Saddle.\\ [2ex]
&&&&Stable node if\\
$A_{2}$&$\frac{s_*^2}{2}-3$&$s_*^2-3(w+1)$&$-s_*\,dg(s_*)$&  $s_*^2<3(w+1)$, $s_*dg(s_*)>0$\\
&&&&Saddle node otherwise \\ [3ex]
&&&&Stable node if \\
$A_{3}$&$\frac{3}{2}(w-1)$&$\frac{3}{2}(w+1)+\Delta s_*\xi$&$2 \Delta \xi dg(s_*)$& $s_* \xi<-\frac{3}{2}(w+1)\sqrt{(1-w)^2+\frac{2\xi^2}{3}}$,\\
&&&& $\xi dg(s_*)<0$ \\
&&&& Saddle node otherwise \\ [2ex]
\multirow{2}{*}{$A_{4}$} & \multirow{2}{*}{$-3(w+1)$} & \multirow{2}{*}{$-\frac{3}{2}\left(1+ \sqrt{1-\frac{4}{3} \,g(0)}\right)$} & \multirow{2}{*}{$-\frac{3}{2}\left(1- \sqrt{1-\frac{4}{3} \,g(0)}\right)$} & Stable if $g(0)>0$ \\
&&&& Saddle if $g(0)<0$\\ [1ex]

\hline\hline
\end{tabular}
\end{adjustbox}
\caption{Stability of critical points of $(0,0)$ model (Sec.~\ref{subsec:00model}).}
\label{tab:eigen_der_A}
\end{table}

The critical points of the system (\ref{x_der})-(\ref{s_der}) for this particular model are given in Table \ref{tab:der_A} and their corresponding eigenvalues along with their stability criteria are given in Table \ref{tab:eigen_der_A}.
In what follows, $s_*$ represents the solution of equation $g(s)=0$ and $dg(s_*)$ is the derivative of $g$ at $s=s_*$.
The system has five critical points depending on $s_*$ and $\xi$.
Critical points $A_{1\pm}$, $A_2$, $A_3$ depend on the concrete form of the scalar field potential through $s_*$, whereas critical point $A_4$ corresponds to the case where the potential is effectively constant.
Note that critical point $A_2$ reduces to point $A_4$ when $s_*=0$.
The properties of the critical points are the following:

\begin{itemize}
\item  Points $A_{1\pm}$ exist for any values of $s_*$ and $\xi$. They correspond to a decelerated solution dominated by the kinetic energy of the scalar field, with stiff fluid effective EoS ($w_{\rm eff}=1$). Points $A_{1 \pm}$ are unstable node whenever $ \pm s_*<\sqrt{6}$ and $\pm dg(s_*)<0$, otherwise they are saddle.

\item Point $A_{2}$ exists for $s_*^2\leq 6$. It corresponds to a scalar field dominated solution. It also corresponds to an accelerated universe when $s_*^2<2$. It is a stable node if $s_*^2<3(w+1)$ and $s_*dg(s_*)>0$, otherwise it is a saddle.

\item  Point $A_3$ exists for any values of $\xi$ and $s_*$. It corresponds to a decelerated solution where the energy content of the universe is shared between matter and the kinetic energy of the scalar field. It is a stable node if  $s_* \xi<-\frac{3}{2}(w+1)\sqrt{(1-w)^2+\frac{2\xi^2}{3}}$ and $\xi dg(s_*)<0$, otherwise it is a saddle.

\item Point $A_4$ corresponds to an accelerated, scalar field dominated solution with $w_{\rm eff}=-1$. It is a stable node whenever $0<g(0)<\frac{3}{4}$, it is stable spiral when $g(0)>\frac{3}{4}$ and saddle whenever $g(0)<0$. For $g(0)=0$ linear theory fails to determine its stability, and other mathematical tools, as e.g.~center manifold theory, or numerical techniques should be employed. In these cases the stability of point $A_4$ can be determined numerically once a specific potential has been chosen. For example we have numerically checked that for some phenomenologically interesting potentials, such as for example $V=\frac{M^{4+n}}{\phi^n}$ (with $M$ and $n>0$ constants), point $A_4$ is stable.
\end{itemize}

\begin{figure}
\centering
\includegraphics[width=8cm,height=5.5cm]{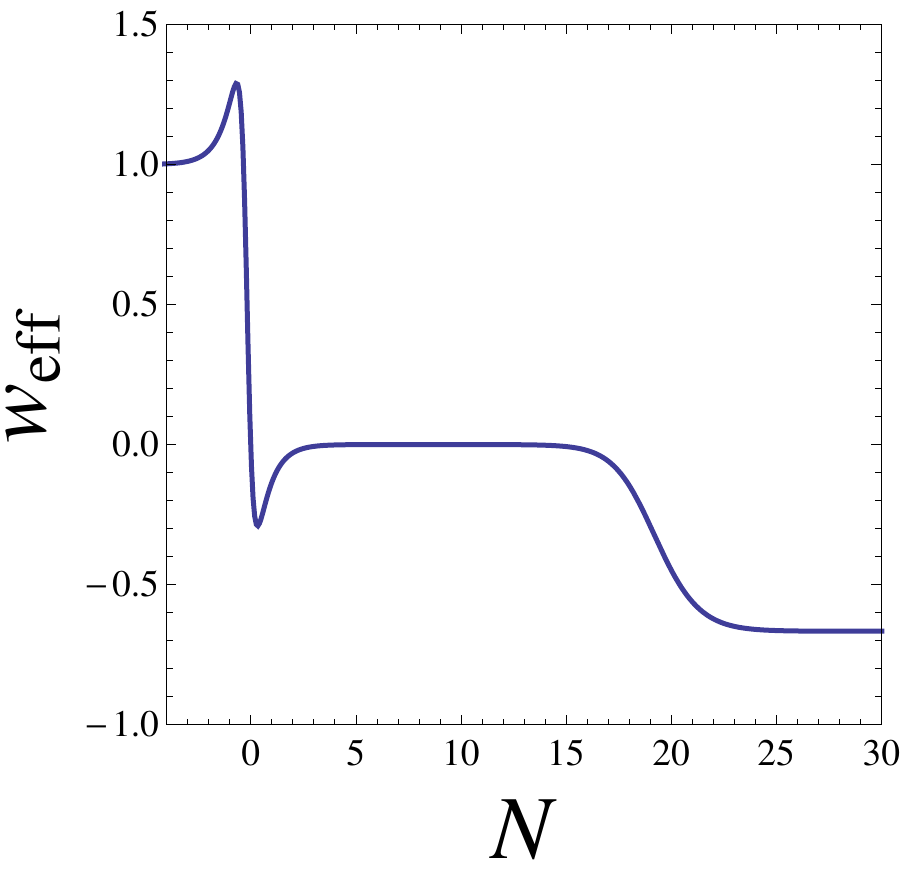}
\caption{Plot of $w_{\rm eff}$ versus $N$ for $(0,0)$ model. Here we have considered the potential $V=V_0\,\sinh^{-\eta}(\mu\phi)$ with $w=0$, $\xi=-1.5$, $\eta=1$ and $\mu=1$.} \label{fig:weff_sinh_I}
\end{figure}

From this analysis, we observe that depending on the choice of the scalar field potential, on the choice of the parameter $\xi$, as well as on the initial conditions, the universe evolves from a stiff matter dominated solution (points $A_{1\pm}$) either towards a decelerated scaling solution (point $A_3$), or towards an accelerated, scalar field dominated solution (points $A_2$ or $A_4$).
The standard matter dominated solution of the canonical scalar field model is replaced by point $A_3$, which is not a matter dominated solution but has $w_{\rm eff}=0$ for $w=0$, i.e.~it behaves as if the universe was matter dominated even though $\Omega_m \neq 1$. 
This feature appears also for canonical scalar field DE models coupled to the matter sector \cite{Amendola:1999er}.
Moreover, point $A_3$ can become the late time attractor, unlike the matter dominated solution of the canonical scalar field model.
This model can thus be used to describe the late time transition of our universe from a DM effective behavior (point $A_3$) to DE domination (points $A_2$ or $A_4$).
Note that point $A_2$ can be a late time accelerating scaling solution (see Fig.~\ref{fig:weff_sinh_I} for an example), which can be used to alleviate the cosmic coincidence problem.

 %%%%%%%%%%%%%%%%%%%%%%%%%%%%%%%%%%%%%%%%%%%%%%%%%%%%%%%%%%%%%%%%%%%%%%%%%%%%%%%%%%%%%%%%%%%%%%%%%%%%%%%%%%%%%%%%%%%%%%

\subsubsection{$(1,0)$ model}
\label{subsec:10model}

\begin{table}%[!ht]
\centering
Here: $\Delta=\sqrt{{\xi}^{2}\,{s_*}^{2} \left( \left( {\xi}^{2}+4 \right) {s_*}^{2}-12(w+1) \right)} $, $x_2=\sqrt{\frac{2}{2\xi^2\,s_*^2+3(1-w)^2}}$, $ \Xi= \frac{1}{2}\frac {\xi\,\sqrt {2\,{s_*}^{2} \left( {\xi}^{2}+2 \right) -
2\,\Delta-12\,w-12}s_*\, \left( w+1 \right) +{s_*}^{2}{\xi}^{2} \left( w+1 \right) +2\,{s_*}^{2}w-\Delta \left( w+1 \right) }{{s_*}^{2}}$
\begin{tabular}{cccccc}
  \hline\hline
  Point~ & ~~~~$x$~~~~ & ~~~~$y$ ~~~~&~~~~$s$~~~~&Existence&~~~~$w_{\rm eff}$\\%~~&~~Acceleration\\
  \hline
  $C_0$ & $0$ & $0$ &$0$ &Always&$w$\\[2.5ex]%&No\\
   $C_{1\pm}$ & $\pm 1$ & 0 &$s_*$ &Always&1\\[2.5ex]%&No\\
   $C_2$&$x_2$&$0$&$s_*$&Always&$w$\\[2.5ex]
    $C_{3}$ & $\frac{1}{2}\frac{\sqrt{6}(w+1)}{s_*}$ & $\frac {\sqrt {3(1-w^2)-{s_*}^{2}{\xi}^{2}+\Delta}}{\sqrt{2}\,s_*}$ &$s_*$ &Fig. \ref{fig:stability_reg_c3_c4}&$\Xi$\\[2.5ex]%&No\\
   $C_4$ & $\frac{s_*}{\sqrt{6}}$ & $\sqrt{1-\frac{s_*^2}{6}}$ &$s_*$ &$s_*^2<6$&$-1+\frac{s_*^2}{3}$\\[1.5ex]
   $C_5$ & $0$ & $1$ &$0$ &Always&$-1$\\[1.5ex]%&No\\
  \hline\hline
\end{tabular}
\caption{Critical points of $(1,0)$ model (Sec.~\ref{subsec:10model}).}
\label{tab:der_1_0}
\end{table}

\begin{table}%[!ht]
\centering
Here: $\eta_\pm=-3+\frac{3}{4}s_*^{2}-\frac{3}{2}w \pm \frac{1}{4}\sqrt {-4\,{s_*}^{4}{\xi}^{2}+{
s_*}^{4}+24\,{s_*}^{2}{\xi}^{2}-12\,{s_*}^{2}w+36\,{w}^{2}}$
\begin{adjustbox}{width=1\textwidth}
\small
\begin{tabular}{ccccc}
\hline\hline
\mbox{Point} & $\lambda_1$  &  $\lambda_2$  &$\lambda_3$  &  Stability \\
\hline  \\[0.1ex]
$C_0$&$\frac{3}{2}(1+w)$&$-\frac{3(1-w)}{4}\Big(1-\sqrt{\frac{16}{3}\frac{g(0)\xi}{(1-w)}+1}\Big)$&$-\frac{3(1-w)}{4}\Big(1+\sqrt{\frac{16}{3}\frac{g(0)\xi}{(1-w)}+1}\Big)$&Saddle\\ [2.5ex]
$C_{1\pm}$&$3(1-w)$&$3 \mp \sqrt{\frac{3}{2}}\,s_*$&$\mp \sqrt{6}\,dg(s_*)$& Unstable node/ Saddle\\[3ex]
$C_{2}$&$-\frac{3}{2}(1-w)$&$\frac{3}{2}(w+1)-\sqrt{\frac{3}{2}}s_*^2\,\xi\,x_2$&$-\sqrt{6}\,x_2\,\xi\,s_*\,dg(s_*)$& Stable node/Saddle\\[2ex]
$C_{3}$& - & - &$-\frac{3(w+1)\,dg(s_*)}{s_*}$&Fig. \ref{fig:stability_reg_c3_c4}\\[2ex]
$C_{4}$& $\eta_+$&$\eta_-$&$-s_*\,dg(s_*)$&Fig. \ref{fig:stability_reg_c3_c4} \\[2.5ex]
&&&&Saddle if $g(0)<0$\\
$C_{5}$& $-3(w+1)$&$-\frac{3}{2}\Big(1+\sqrt{1-\frac{4\,g(0)}{3}}\Big)$&$-\frac{3}{2}\Big(1-\sqrt{1-\frac{4\,g(0)}{3}}\Big)$&Stable if $g(0)>0$ \\
&&&&See Appendix~\ref{App A} if $g(0)=0$\\
\hline\hline
\end{tabular}
\end{adjustbox}
\caption{Stability of critical points listed in Table \ref{tab:der_1_0}. The expressions for the eigenvalues of point $C_3$ have not been written due to their excessive length.}
\label{tab:eigen_1_0}
\end{table}

\begin{figure}
\centering
\includegraphics[width=8cm,height=5.5cm]{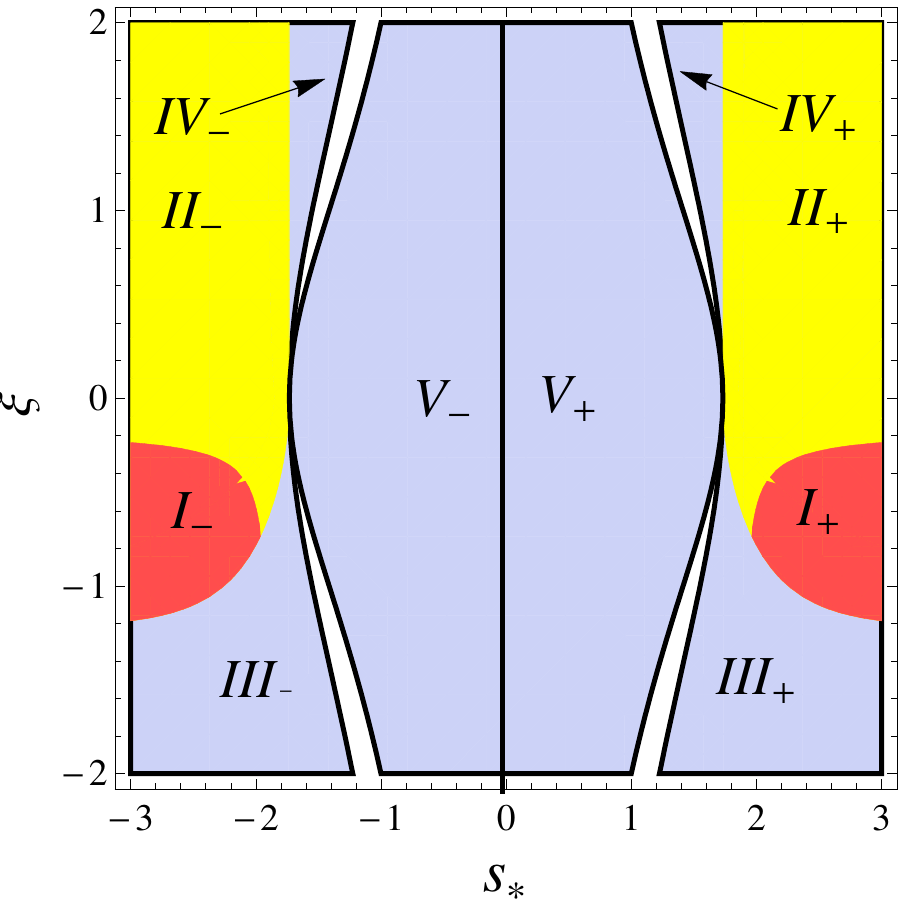}%}
\caption{Existence and stability regions of points $C_3$, $C_4$ on the ($s_*$, $\xi$) parameter space. Regions $I_+$, $I_-$, $II_+$, $II_-$, $III_+$, $III_-$, $IV_+$, $IV_-$ represent regions of existence of point $C_3$. Regions $I_+$ and $II_+$ represent regions of stability of point $C_3$ for potentials where $dg(s_*)>0$ and regions $I_-$ and $II_-$ represent its regions of stability for potentials where $dg(s_*)<0$. Red shaded regions (i.e.~regions $I_+$ and $I_-$) represent regions of acceleration for point $C_3$. Region $V_+$ represents the region of stability of point $C_4$ for potential with $dg(s_*)>0$ and region $V_-$ represents its region of stability for potential with $dg(s_*)<0$. Here we have taken $w=0$.}
\label{fig:stability_reg_c3_c4}
\end{figure}

This section deals with the phase space analysis of the dynamical system~(\ref{x_der})-(\ref{s_der}) for the choices of $\beta=1$ and $\alpha=0$.
In terms of dimensionless variables (\ref{variable1}), the effective EoS parameter $w_{\rm eff}$ is given by
\begin{align}
  w_{\rm eff} &=x^2-y^2+w(1-x^2-y^2)+\sqrt{\frac{2}{3}}\, x\,\xi\, s\,(1-x^2-y^2)^{1/2} \,.
\end{align}
The critical points of the system (\ref{x_der})-(\ref{s_der}) for this model are given in Table \ref{tab:der_1_0} and their corresponding eigenvalues along with their stability criteria are given in Table \ref{tab:eigen_1_0}.
The system has six critical points depending on the values of $s_*$ and $\xi$.
All critical points depend on the concrete form of the potential $V(\phi)$ through $s_*$.
Points $C_0$ and $C_5$ correspond to the case where the variable $s=0$, i.e.~when the potential is effectively constant.
The properties of the critical points are as follow:
\begin{itemize}

\item  Point $C_0$ corresponds to a matter dominated solution ($\Omega_m = 1$) with effective EoS parameter coinciding with the matter one ($w_{\rm eff}=w$). It is always saddle.

\item  Points $C_{1\pm}$ exist for any values of $s_*$ and $\xi$. They correspond to a decelerated solution dominated by the kinetic energy of the scalar field, with stiff fluid effective EoS ($w_{\rm eff}=1$). Point $C_{1\pm}$ are unstable nodes whenever $\pm s_*<\sqrt{6}$ and $\pm dg(s_*)<0$, otherwise they are saddle.

\item Point $C_{2}$ corresponds to a decelerated scaling solutions with effective EoS mimicking a matter era: $w_{\rm eff}=w$. It is stable when $\xi>\sqrt{\frac{3}{2}}\frac{(w+1)}{s_*^2\,x_2}$ and $\xi\,s_*dg(s_*)>0$, otherwise it is saddle.

\item  Due to the complicated expressions of point $C_3$, we numerically determine its regions of existence and stability in the $(s_*,\xi)$ parameter space assuming $w=0$ (see Fig.~\ref{fig:stability_reg_c3_c4}). In the same figure we report the regions of parameter space where point $C_3$ can describe a late time accelerated scaling solution.

\item Point $C_{4}$ corresponds to a scalar field dominated solution. It exists whenever $s_*^2<6$ and describes an accelerated universe if $s_*^2<2$. It reduces to the critical point $C_5$ for $s_*=0$. The stability conditions of this point are again determined numerically by plotting the regions of stability in the $(s_*,\xi)$ parameter space (see Fig. \ref{fig:stability_reg_c3_c4}).

\item Point $C_{5}$ corresponds to an accelerated scalar field dominated solution with effective EoS $w_{\rm eff}=-1$. It is is a saddle if $ g(0)<0$, while it represents a late time attractor if $g(0)>0$. If $g(0)=0$, linear stability fails. In this case we use center manifold theory to determine its stability. The detailed analysis using these advanced tools is presented in appendix \ref{App A}. From that analysis, we find that point $C_5$ is always unstable unless $dg(0)=0$.

\end{itemize}

\begin{figure}
\centering
\includegraphics[width=8cm,height=5.5cm]{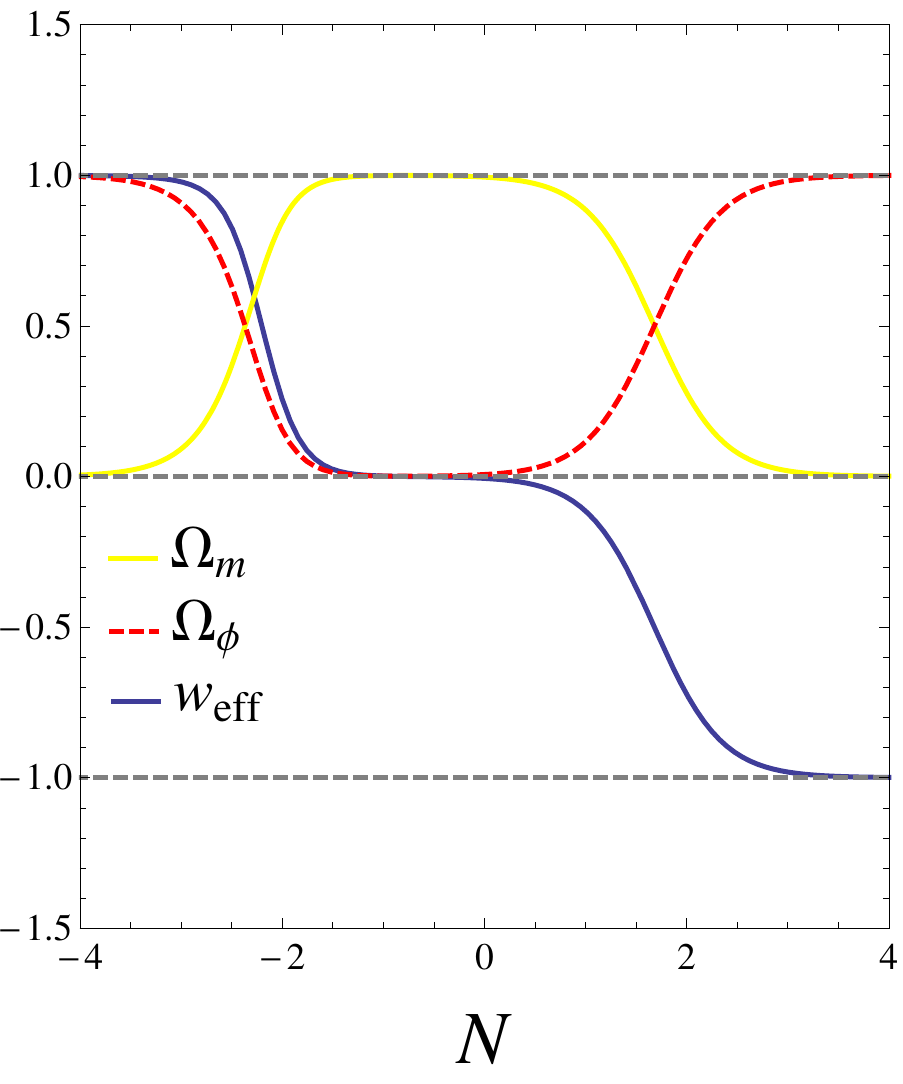}
\caption{Plot of $\Omega_m$, $\Omega_\phi$ and $w_{\rm eff}$ versus $N$ for  $(1,0)$ model. Here we have considered the potential $V=V_0\,\sinh^{-\eta}(\mu\phi)$ with $w=0$, $\alpha=0$, $\xi=-1.5$, $\eta=-1$ and $\mu=1$.} \label{fig:weff_sinh_II}
\end{figure}

From the analysis of the critical points, we understand that, depending on the choice of the potential $V(\phi)$, of the parameter $\xi$, as well as the initial conditions, the universe can evolve from a stiff matter dominated solution (points $C_{1\pm}$) either towards an accelerated scaling solution $C_3$, or towards an accelerated, scalar field dominated solution (points $C_4$ or $C_5$), passing through a long lasting matter dominated solution (point $C_0$) or a decelerated scaling solution (point $C_2$) with $w_{\rm eff}=w$.
This means that the $(1,0)$ model can be used to describe the observed transition of the universe from a matter dominated era to a late time DE dominated era (see Fig.~\ref{fig:weff_sinh_II} for an example).
Moreover the background dynamics of this model presents different scaling solutions (points $C_2$, $C_3$ and $C_4$) which can be used to obtain interesting phenomenology: for example late time accelerated scaling solutions can be used to alleviate the cosmic coincidence problem and scaling solutions mimicking a matter era could present interesting observational signatures at the perturbation level without affecting the background dynamics (see Sec.~\ref{sec:perturbation}).

\subsubsection{$(1,\frac{1}{2})$ model}
\label{subsec:112model}

\begin{table}%[!ht]
  \centering
  \tiny
Here: $\Delta=\sqrt{\xi^2\,s_*^2+6(1-w^2)}$, $ \Xi_4= \frac {2\,{\xi}^{2}{s_*}^{4}-3\,{\xi}^{2}{s_*}^{2}-2\,{s_*}^{4}+
12\,{s_*}^{2}-18}{2\,{\xi}^{2}{s_*}^{4}}
$\\
\begin{adjustbox}{width=1\textwidth}
\small
\begin{tabular}{cccccc}
  \hline\hline
  Point~ & ~~~~$x$~~~~ & ~~~~$y$ ~~~~&~~~~$s$~~~~&Existence&~~~~$w_{\rm eff}$\\%~~&~~Acceleration\\
  \hline
  $D_0$ & $0$ & 0 &$s$ &Always&$w$\\[1.5ex]%&No\\
   $D_{1\pm}$ & $\pm 1$ & 0 &$s_*$ &Always&1\\[1.5ex]%&No\\
    $D_{2\pm}$ & $\frac{1}{2}\frac{\sqrt{6}(w+1)}{s_*}$ & $\frac{\xi s_*\pm \Delta}{2\,s_*}$ &$s_*$ &$0\leq (\xi s_* \pm \Delta)^2+6(w+1)\leq 4\,s_*^2$&$w$\\[1.5ex]%&No\\
   $D_3$ & $\frac {s_* \left(\xi \sqrt {{\xi}^{2}{s_*}^{2}-{s_*}^{2}+6}+\sqrt {6} \right) }{({\xi}^{2}{s_*}^{2}+6)}
$ & $-{\frac {\sqrt {6\,({\xi}^{2}{s}^{2}-{s}^{2}+6)}-\xi\,{s}^{2}}{{\xi}
^{2}{s}^{2}+6}}$ &$s_*$ &Always&$-\frac {-\xi\,\sqrt {{\xi}^{2}{s_*}^{2}-{s_*}^{2}+6}\sqrt {6}{s_*}^{2}
+3\,({\xi}^{2}{s_*}^{2}-2\,{s_*}^{2}+6)}{3({\xi}^{2}{s}^{2}+6)}
$\\[1.5ex]
 $D_4$ & $\frac{\sqrt{6}}{2\,s_*}$ & $\frac{s_*^2-3}{\xi\,s_*^2}$ &$s_*$ &$0 \leq \frac {3\,{\xi}^{2}{s_*}^{2}+2\,{s_*}^{4}-12\,{s_*}^{2}+18}{2\,{\xi}^{2}{s_*}^{4}}
 \leq 1$&$\Xi_4$\\[1.5ex]
   $D_{5\pm}$ & $0$ & $\pm 1$ &$0$ &Always&$-1$\\[1.5ex]
  \hline\hline
\end{tabular}
\end{adjustbox}
\caption{Critical points of $(1,\frac{1}{2})$ model (Sec.~\ref{subsec:112model}).}
\label{tab:der_1_half}
\end{table}

\begin{table}%[!ht]
\centering
\tiny
Here: $\xi_{1\pm}=-\frac{1}{4}(1-w)\pm\frac{1}{3s_*}\left(12\,{\xi}^{4}{s_*}^{4}+ 12\,\Delta\,\xi\,s_*\left({
\xi}^{2}{s_*}^{2}+{s_*}^{2} + 3{w}^{2}+3 w+ 6\right)-12\,{\xi}^{2}{s_*}^{4}-72\,{\xi}^{2}{s_*}^{2}{w}^
{2}\right.$\\$\left.+36\,{
\xi}^{2}{s_*}^{2}w+108\,{\xi}^{2}{s_*}^{2}+81
\,{s_*}^{2}{w}^{2}-18\,w{s_*}^{2}-63\,{
s_*}^{2}+216\,(w+1)(1-w^2)\right)^\frac{1}{2}$\\
$\xi_{2\pm}=-\frac{1}{4}(1-w)\pm\frac{1}{3s_*}\left(12\,{\xi}^{4}{s_*}^{4}+ 12\,\Delta\,\xi\,s_*\left({
\xi}^{2}{s_*}^{2}+{s_*}^{2} + 3{w}^{2}+3 w+ 6\right)-12\,{\xi}^{2}{s_*}^{4}-72\,{\xi}^{2}{s_*}^{2}{w}^
{2}\right.$\\$\left.+36\,{
\xi}^{2}{s_*}^{2}w+108\,{\xi}^{2}{s_*}^{2}+81
\,{s_*}^{2}{w}^{2}-18\,w{s_*}^{2}-63\,{
s_*}^{2}+216\,(w+1)(1-w^2)\right)^\frac{1}{2}$\\

\begin{adjustbox}{width=1\textwidth}
\small
\begin{tabular}{ccccc}
\hline\hline
\mbox{Point} &~~~~~~~~~~~~$\lambda_1$~~~~~~~~~ &~~~~~~~~~~~~ $\lambda_2$~~~~~~~~~~~~~~~~~~~~ &$\lambda_3$ ~~~~~~~~~~~~~~~~~& ~~~~~~~~~Stability \\
\hline  \\[0.1ex]
$D_0$&$0$&$-\frac{3}{2}(1-w)$&$\frac{3}{2}(1+w)$&Saddle.\\ [2ex]
$D_{1\pm}$&$3(1-w)$&$3 \mp \sqrt{\frac{3}{2}}\,s_*$&$\mp \sqrt{6}\,dg(s_*)$& Unstable node/ Saddle\\[3ex]
$D_{2\pm}$&$\xi_{1\pm}$&$\xi_{2\pm}$&$-\frac{3(w+1)}{s_*} dg(s_*)$& Fig.~\ref{fig:stability_reg_D2} \\[3ex]
$D_{3}$&$-\frac {-\xi\,\sqrt {6({\xi}^{2}{s_*}^{2}-{s_*}^{2}+6)}\,{s_*}^{
2}+3\,({\xi}^{2}{s_*}^{2}-2\,{s_*}^{2}+6)}{{\xi}^{2}{s_*}^{2}+6}$&$-\frac {-\xi\,\sqrt {6({\xi}^{2}{s_*}^{2}-\,{s}^{2}+6)}\,{s_*}^{2}+6\,({\xi}^{2}{s_*}^{2}-{s_*}^{2}+6)}{2({\xi}^{2}{s_*}^{2}+6)}
$&$-{\frac {\sqrt {3}s_* \left( \xi\,\sqrt {2\,{\xi}^{2}{s_*}^{2}-2\,{s_*}^{2}+
12}+2\,\sqrt {3} \right)\, dg(s_*)}{{\xi}^{2}{s_*}^{2}+6}}
$&Fig. \ref{fig:stability_reg_D3}\\[3ex]
$D_{4}$&$\frac {3(2\,{\xi}^{2}{s*}^{4}-2\,{s_*}^{4}-3\,{\xi}^{2}{s_*}^{2}+12\,{s_*}^{2}-18)}{4\,{\xi}^{2}{s*}^{4}}
$&$\frac {3(4\,{\xi}^{2}{s*}^{4}-6\,{s_*}^{4}-9\,{\xi}^{2}{s_*}^{2}+36\,{s_*}^{2}-54)}{4\,{\xi}^{2}{s*}^{4}}
$&$-\frac{3 dg(s_*)}{s_*}$&Unstable node/ Saddle\\[3ex]
&&&& Saddle if $ g(0)(1 \mp \xi)<0$ \\
$D_{5\pm}$&$-3(w+1)$&$-\frac{3}{2}+\frac{1}{2}\sqrt{9-12 g(0)(1 \mp \xi)}$&$-\frac{3}{2}-\frac{1}{2}\sqrt{9-12 g(0)(1 \mp \xi)}$& Stable if $g(0)(1 \mp \xi)>0$ \\
&&&& See Appendix~\ref{App A} if $  g(0)(1 \mp \xi)=0$ \\[1ex]
\hline\hline
\end{tabular}
\end{adjustbox}
\caption{Stability of critical points listed in Table \ref{tab:der_1_half}. Points $D_3$ and $D_4$ are analysed only for $w=0$.}
\label{tab:eigen_der_1_half}
\end{table}

\begin{figure}
\centering
\subfigure[]{%
\includegraphics[width=6cm,height=4cm]{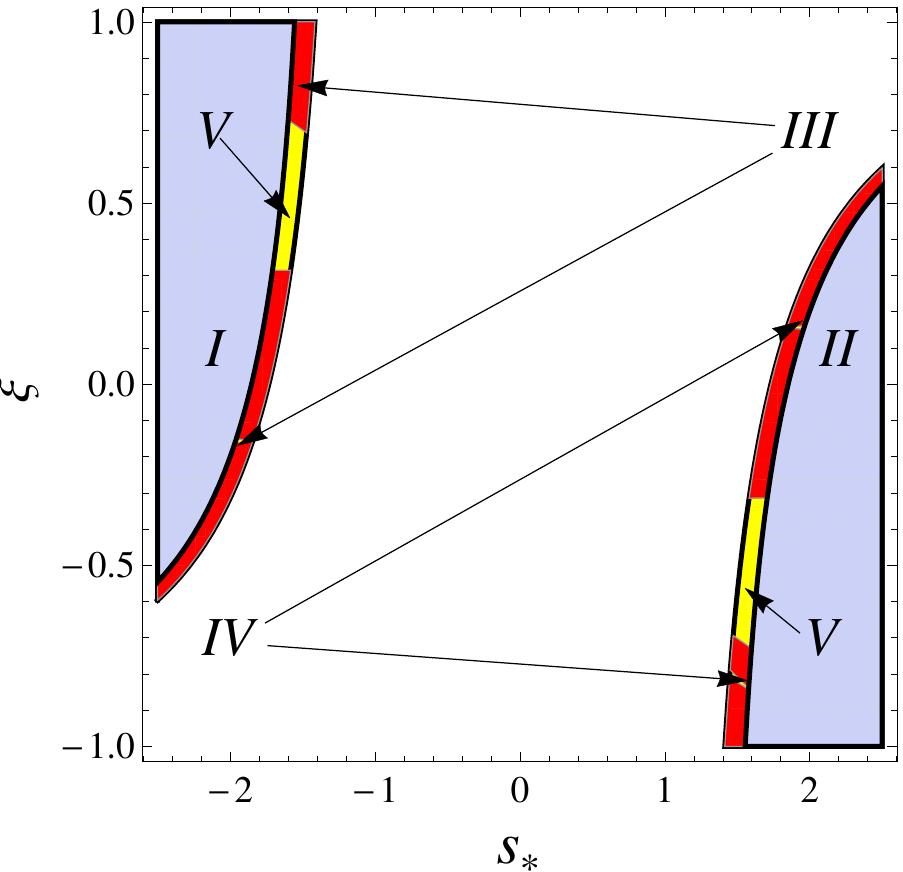}\label{fig:stability_reg_D2p}}
\qquad
\subfigure[]{%
\includegraphics[width=6cm,height=4cm]{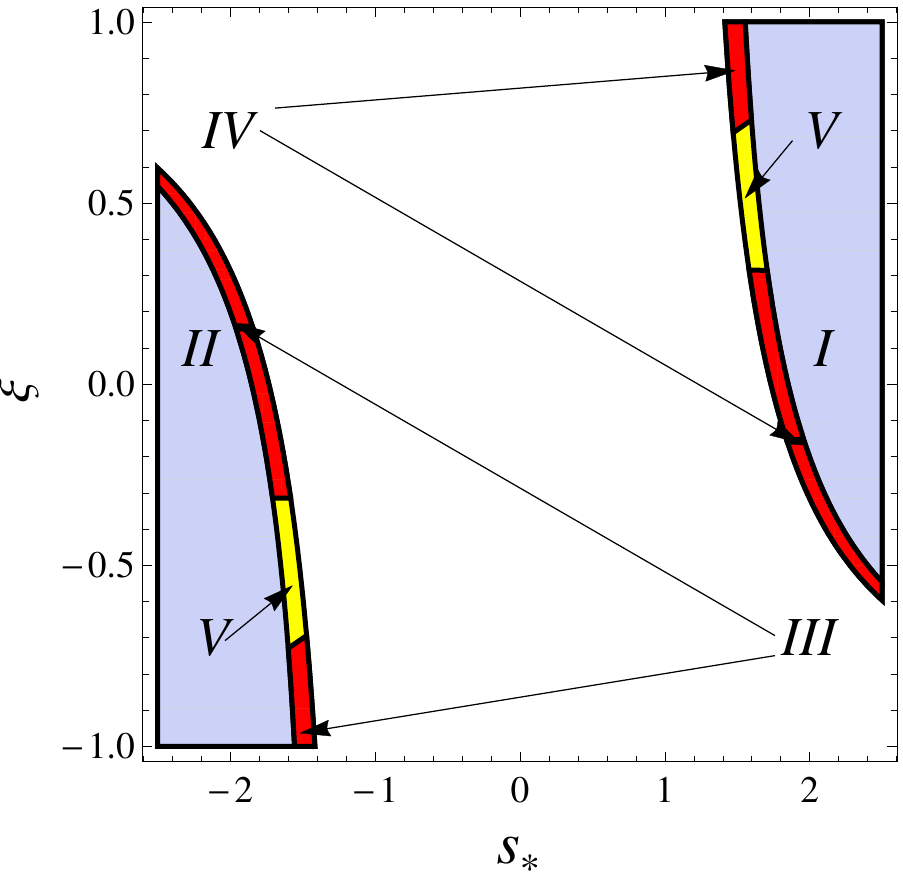}\label{fig:stability_reg_D2n}}
\caption{Existence and stability regions of point $D_{2+}$ (a) and point $D_{2-}$ (b) on the ($s_*$, $\xi$) parameter space. In both panels, the whole shaded regions represent regions of existence. Regions $I$ and $II$ represent regions where the point is a stable spiral for potentials giving $dg(s_*)<0$ and $dg(s_*)>0$, whereas regions $III$ and $IV$ represent regions where the point is a stable node for potentials giving $dg(s_*)<0$ and $dg(s_*)>0$ respectively. Yellow shaded region $V$ represents the region where the point is saddle. Here we have assumed $w=0$.
}
\label{fig:stability_reg_D2}
\end{figure}

\begin{figure}
\centering
\includegraphics[width=8cm,height=5.5cm]{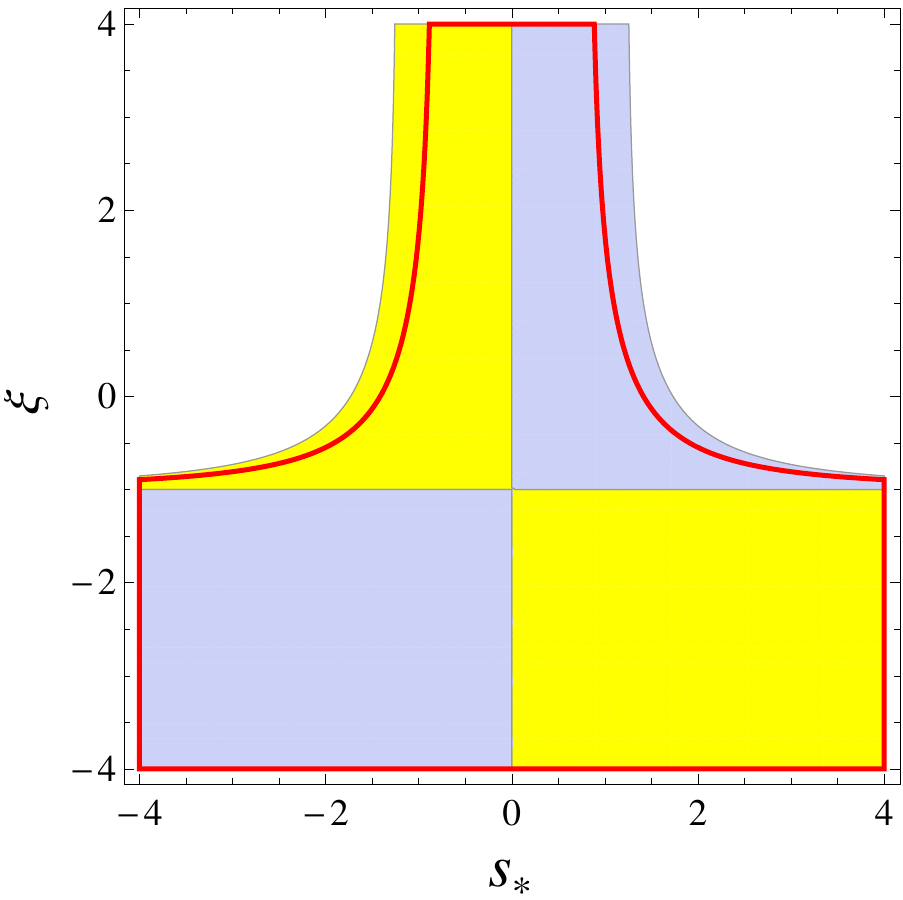}%\label{fig:stability_reg_D3}
\caption{Existence and stability regions of point $D_{3}$ on the ($s_*$, $\xi$) parameter space. The yellow shaded regions  represent regions of stability of the point for potentials giving $dg(s_*)<0$, whereas blue shaded regions  represent regions of stability of the point for potentials giving $dg(s_*)>0$. Region enclosed inside the red colored boundary corresponds to the region of acceleration.}
\label{fig:stability_reg_D3}
\end{figure}

This section deals with the phase space analysis of the dynamical system~(\ref{x_der})-(\ref{s_der}) for $\beta=1$ and $\alpha=\frac{1}{2}$.
In terms of the dimensionless variables (\ref{variable1}), the effective EoS parameter $w_{\rm eff}$ is given by
\begin{align}
  w_{\rm eff} \equiv\frac{p+\frac{1}{2}\dot{\phi}^2-V+p_{\rm int}}{\rho+\frac{1}{2}\dot{\phi}^2+V+\rho_{\rm int}}%\\
   = x^2-y^2+w(1-x^2-y^2)+\sqrt{\frac{2}{3}}\xi\, s \, x\,y
\end{align}
The critical points of the system (\ref{x_der})-(\ref{s_der}) are given in Table \ref{tab:der_1_half} and their corresponding eigenvalues along with their stability criteria are given in Table \ref{tab:eigen_der_1_half}.
The system presents nine critical points depending on $s_*$ and $\xi$.
Note that critical point $D_3$ reduces to point $D_{5-}$ when $s_*=0$.
The properties of these critical points are the following:
\begin{itemize}
\item  Point $D_{0}$ is independent of the specific scalar field potential under considerations for its existence. It corresponds to a matter dominated solution with $w_{\rm eff}=w$. It always behaves as a saddle.
\item  Points $D_{1\pm}$ exist for any values of $s_*$ and $\xi$. They correspond to a decelerated solution dominated by the kinetic energy of the scalar field, with stiff fluid effective EoS ($w_{\rm eff}=1$). Points $D_{1 \pm}$ are unstable node whenever $\pm s_*<\sqrt{6}$ and $\pm dg(s_*)<0$, otherwise they are saddle.
\item Points $D_{2\pm}$ correspond to decelerated scaling solutions with effective EoS mimicking a matter era ($w_{\rm eff}=w$). The stability condition for points $D_{2\pm}$ cannot be determined analytically due to the complicated expressions of their eigenvalues. However they can be stable for some values of $s_*$ and $\xi$, as checked numerically and shown in Fig.~\ref{fig:stability_reg_D2} by plotting their regions of existence and stability in the $(s_*,\xi)$ parameter space for $w=0$. In any case since they always constitute decelerated solutions, these points cannot describe the late time acceleration of the universe.
\item  Due to the extremely complicated expressions associated to point $D_3$, we are able to determine its stability only fixing the parameter $w$. For this purpose we choose $w=0$. In this case point $D_3$ exists for any values $s_*$ and $\xi$. It corresponds to a late accelerated scalar field dominated solution ($\Omega_\phi=1$) for some values of $s_*$ and $\xi$ (see Fig. \ref{fig:stability_reg_D3}).
\item Point $D_4$ is again analysed only for the $w=0$ case due to the complicated expressions associated to it. It exists for $0 \leq \frac {3\,{\xi}^{2}{s_*}^{2}+2\,{s_*}^{4}-12\,{s_*}^{2}+18}{2\,{\xi}^{2}{s_*}^{4}}
 \leq 1$. Numerically we have checked that this point is not stable within its region of existence: it is either saddle or an unstable node.
\item Points $D_{5\pm}$ correspond to accelerated, scalar field dominated solutions with $w_{\rm eff}=-1$. Point $D_{5-}$ is a special case of point $D_3$ for $s_*=0$.  Point $D_{5+}$ is a saddle if $ g(0)(1- \xi) <0$, it is  stable if $ g(0)(1- \xi) >0$ but linear stability fails to determine the stability if $ g(0)(1 - \xi)=0$. Similarly, point $D_{5-}$ is a saddle if $ g(0)(1+ \xi) <0$, it is  stable if $ g(0)(1+ \xi) >0$  and linear stability fails to determine the stability if $ g(0)(1 + \xi)=0$, in which case other  mathematical tools, as for example center manifold theory, are required to complete the analysis. The complete investigation using center manifold techniques for point $D_{5-}$ and $D_{5+}$ is reported in appendix \ref{App A}.
From its results we find that points $D_{5\pm}$ are always unstable unless $dg(0)\,(1\mp\xi)=0$.
\end{itemize}

\begin{figure}
\centering
\includegraphics[width=8cm,height=5.5cm]{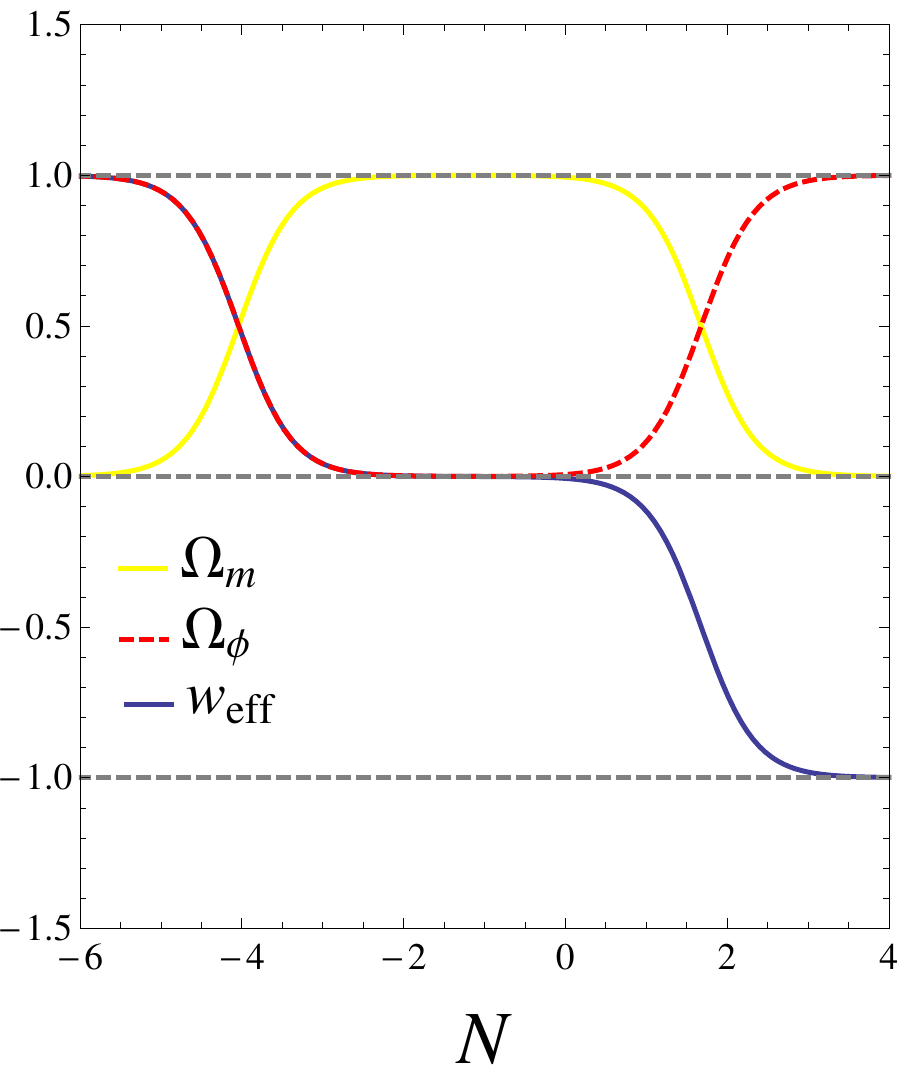}
\caption{Evolution of $\Omega_{\phi}$, $\Omega_{m}$ and $w_{\rm eff}$ versus $N$ for  $(1,\frac{1}{2})$ model. Here we considered a potential $V=V_0\sinh^{-\eta}(\mu\phi)$ with $\eta=-1$, $\mu=1$ $w=0$ and $\xi=1$.}
\label{fig:weff_sinh_III}
\end{figure}

From the analysis of the critical points, it can be observed that the universe evolves from an early-time stiff matter dominated solutions $D_{1\pm}$ towards an accelerated, scalar field dominated solution (points $D_3$, $D_{5\pm}$), possibly passing through a long lasting matter dominated solution $D_0$ (or $D_{2\pm}$).
This model can thus used to phenomenologically describe the observed transition of the universe from a matter dominated era to a late time DE dominated era (see Fig.~\ref{fig:weff_sinh_III} for an example).
Moreover points $D_{2\pm}$ describe scaling solutions which can be used to characterize a matter era with $w_{\rm eff} = w$.
Since for these solutions the energy density of the scalar field does not vanish, some deviations from standard $\Lambda$CDM dynamics might be present at the perturbation level, even if the background evolution is undistinguishable (see Sec.~\ref{sec:perturbation}).

%%%%%%%%%%%%%%%%%%%%%%%%%%%%%%%%%%%%%%%%%%%%%%%%%%%%%%%%%%%%%%%%%%%%%%%%%%%%%%%%%%%%%%%%%%%%%%%%

\subsection{Model II}
\label{sec:model2}

In this section we present the phase space analysis of the system~(\ref{x_der})-(\ref{s_der}) for Model~II (see Table~\ref{models}), where the coupling function $f$ is of the form $\xi \Big(-\frac{1}{V}\frac{dV}{d\phi}\Big)^\beta\,\frac{H_0}{\mathfrak{\kappa\,n}}$,
with $\xi$ a constant and $H_0$ the Hubble constant.
It can be seen from Table \ref{models} that the quantity $A$ does not depend solely on $x$, $y$, $s$ and hence, as discussed previously, another extra variable $z$ is required to close the autonomous system~(\ref{x_der})-(\ref{s_der}).
In what follows we will choose this further variable as
\begin{align}
z=\frac{H_0}{H+H_0} \,,
\end{align}
which is bounded as $0\leq z \leq 1$.
The dynamical system~(\ref{x_der})-(\ref{s_der}) becomes then
\begin{align}
x'&=-\frac{1}{2(z-1)}\Big[3 x (z-1) \left(1-w+(w-1)x^2+(1+w)y^2\right)\nonumber\\
& \qquad\qquad\qquad\qquad\qquad +\sqrt{6}\left(-\xi\,s^\beta\,z(1-x^2)-s(z-1)y^2\right)\Big],\label{x_der_B}\\
y'&=-\frac{y}{2(z-1)}\left[3(z-1)\left((1+w)(y^2-1)+(w-1)x^2\right)+\sqrt{6}x\left((z-1)s+z\xi\,s^\beta\right)\right],\label{y_der_B}\\
z'&=\frac{z}{2}\left[3(z-1)\left((1+w)(y^2-1)+(w-1)x^2\right)+\sqrt{6}z \xi x s^\beta \right],\label{z_der_B}\\
s'&=-\sqrt{6}\,x\,g(s)\label{s_der_B}.
\end{align}
The four dimensional phase space of the system (\ref{x_der_B})-(\ref{s_der_B}) is given by
\begin{equation}
\Psi=\left\lbrace(x,y,z)\in \mathbb{R}^3:0\leq x^2+y^2\leq 1, 0 \leq z \leq 1 \right\rbrace \times \left\lbrace s \in \mathbb{R}\right\rbrace.
\end{equation}
The acceleration equation (\ref{acc_eqn_er}) for this coupling model yields
\begin{align}
\frac{\dot{H}}{H^2}=\frac{3}{2}\left\lbrace -(w+1)-(1-w) x^2+(w+1) y^2-\frac{2\,s^\beta\, \xi}{\sqrt{6}}\frac{z x}{(1-z)}\right\rbrace \,.
\end{align}
Hence, the effective EoS parameter is given by
\begin{align}
w_{\rm eff}=w+(1-w) x^2-(w+1) y^2+\frac{2 s^\beta\,\xi}{\sqrt{6}}\frac{z x}{(1-z)}
\end{align}
Here will investigate only the cases $\beta=0$ and $\beta=1$ (cf.~Table~\ref{model_I}), since the complete analysis with a general $\beta$ would be too complicated.

\subsubsection{$\beta=0$ case}
\label{subsec:beta0}

\begin{table}%[!ht]
\centering
\begin{tabular}{ccccccc}
  \hline\hline
  Point ~~&~~ $x$~~~~ &~~~~~~ $y$~~~~~~ & ~~~~$z$~~~~&~~~~$s$~~~~&~~~~Existence~~~~&~~~~$w_{\rm eff}$\\
  \hline
  $E_0$ & $0$ & 0 & 0&$s$&Always&$w$ \\
   $E_{1\pm}$ & $\pm 1$ & $0$ & $0$&$s_*$ &Always&$1$\\
  $E_2$ & $\sqrt{\frac{3}{2}} \frac{(1+w)}{s_*}$ & $\sqrt{\frac{3}{2}} \frac{\sqrt{(1+w)(1-w)}}{s_*}$ & $0$&$s_*$ &Always&$w$\\
  $E_3$ & $\frac{s_*}{\sqrt{6}}$ & $\sqrt{1-\frac{s_*^2}{6}}$ & $0$ &$s_*$&$s_*^2 \leq 6$&$\frac{s_*^2}{3}-1$\\
  $E_4$ & 0 & 1 & $\frac{s_*}{s_*+\xi}$ &$s_*$&$0\leq \frac{s_*}{s_*+\xi}\leq 1$&$-1$\\
  $E_{5\pm}$&$\pm 1$&$0$&$\frac{\sqrt{6}}{\sqrt{6}\mp\xi}$&$s_*$&$\pm \xi<0$&$-1$\\
  $E_6$ & 0 & 1 & $0$ &$0$&Always&$-1$\\
  \hline\hline
\end{tabular}
\caption{Critical points of Model II with $\beta=0$ (Sec.~\ref{subsec:beta0}).}
\label{tab:der_B_0}
\end{table}

\begin{table}%[!ht]
\centering
$\Xi_{\pm}=-\frac{3}{4}(1-w)\pm\frac{3}{4s_*}\sqrt{(1-w)(24(1+w)^2-s_*^2(7+9w))}$\\
\begin{adjustbox}{width=1\textwidth}
\small
\begin{tabular}{cccccc}
\hline\hline
Point &  $\lambda_1$&$\lambda_2$&$\lambda_3$&$\lambda_4$ & Stability \\
\hline\\
$E_0$   &$0$&$\frac{3}{2}(w-1)$&$\frac{3}{2}(w+1)$ &$\frac{3}{2}(w+1)$& Saddle \\[2ex]
$E_{1\pm}$   &$3(1-w)$&$3\mp \frac{\sqrt{6}}{2}s_*$&$3$ &$\mp \sqrt{6}\,dg(s_*)$& Unstable node/saddle \\[2ex]
$E_2$ & $\frac{3}{2}(1+w)$&$\Xi_+$ &$\Xi_-$ &$-\frac{3(w+1)dg(s_*)}{s_*}$& Saddle \\[2ex]
$E_3$ & $\frac{s_*^2-6}{2}$&$\frac{s_*^2}{2}$&   $s_*^2-3(1+w)$&$-s_* dg(s_*)$&Saddle \\[2ex]
$E_4$ & $-3(1+w)$&$-\frac{3}{2}\left(1+ \sqrt{1-\frac{2 s_*^2}{3}}\right)$&$-\frac{3}{2}\left(1- \sqrt{1-\frac{2 s_*^2}{3}}\right)$&$0$ & - \\[2.5ex]
$E_{5\pm}$ & $-3$&$-3(w+1)$&$\mp \frac{\sqrt{6}}{2}s_*$&$\mp \sqrt{6}dg(s_*)$ & Stable node if $\pm s_*>0$, $\pm dg(s_*)>0$ \\
&&&& & Saddle otherwise\\[2ex]
\multirow{2}{*}{$E_6$} & \multirow{2}{*}{$0$} & \multirow{2}{*}{$-3(w+1)$} & \multirow{2}{*}{$-\frac{3}{2}\left(1+ \sqrt{1-\frac{4}{3} \,g(0)}\right)$} & \multirow{2}{*}{$-\frac{3}{2}\left(1- \sqrt{1-\frac{4}{3} \,g(0)}\right)$} & Saddle if $g(0)<0$ \\
&&&& & See Appendix \ref{App A} for $g(0) \geq 0$\\ [1ex]
\hline\hline
\end{tabular}
\end{adjustbox}
\caption{Stability of critical points given in table \ref{tab:der_B_0}.}
\label{tab:eigen_der_B_0}
\end{table}

The critical points of the system (\ref{x_der_B})-(\ref{s_der_B}) for the choice $\beta=0$ are given in Table \ref{tab:der_B_0} and their corresponding eigenvalues along with their stability criteria are given in Table \ref{tab:eigen_der_B_0}.
The system has nine critical points depending on $s_*$ and $\xi$.
Critical points  $E_{1\pm}$, $E_2$, $E_3$, $E_4$ and $E_{5\pm}$ depend on the concrete form of the potentials $V(\phi)$ through $s_*$. Note that critical points $E_3$ and $E_4$ reduce to point $E_6$ when $s_*=0$.
The properties of the critical points are:
\begin{itemize}
\item Point $E_0$ exists for any scalar field potential. It corresponds to a decelerated matter dominated solution with $w_{\rm eff}=w$ and it always behaves as a saddle.
\item Points $E_{1\pm}$ exist for any values of $\xi$ and $s_*$. They correspond to stiff matter solutions ($w_{\rm eff}=1$), dominated by the kinetic part of the scalar field. Point $E_{1 \pm}$ are unstable nodes when $\pm s_*<\sqrt{6}$ and $\pm dg(s_*)<0$, otherwise they are saddle.
\item Point $E_2$ exists for any values of $\xi$ and $s_*$. It corresponds to a decelerated scaling solution with $w_{\rm eff}=w$. It is always saddle since the eigenvalues of the Jacobian matrix always satisfy $\lambda_1>0$ and $\lambda_3<0$.
\item Point $E_3$ exists for $s_*^2\leq 6$. It always corresponds to a scalar field dominated universe and describes an accelerated universe if $s_*^2<2$. It is a saddle for any values of $\xi$ and $s_*$ since at least two of its corresponding eigenvalues have opposite sign: $\lambda_1<0$ and $\lambda_2>0$.
\item Point $E_4$ exists when $0\leq \frac{s_*}{s_*+ \xi} \leq 1$ and it characterizes a universe dominated by the scalar field potential energy, and consequently $w_{\rm eff}=-1$. Since it is a non-hyperbolic point ($\lambda_4=0$), we cannot determine its stability properties using linear stability theory. This point has to be analysed using center manifold theory or numerical techniques only once a specific potential has been selected. We have anyway checked that for some phenomenologically relevant scalar field potentials this point can be stable.
\item Points $E_{5\pm}$ exist when $\mp \xi<0$. They correspond to an accelerated universe mimicking a cosmological constant EoS ($w_{\rm eff}=-1$), and are dominated by the scalar field kinetic energy. Point $E_{5+}$ is a stable node when $s_*>0$ and $dg(s_*)>0$, whereas point $E_{5-}$ is a stable node when $s_*<0$ and $dg(s_*)<0$.
\item Point $E_6$ corresponds to the case where the potential $V(\phi)$ is effectively constant, and thus it describes an accelerated solution dominated by the scalar field potential energy ($w_{\rm eff}=-1$). It behaves as a saddle if $g(0)<0$, while for $g(0)\geq 0$ one cannot use linear stability theory but center manifold theory should be employed. The analysis using this advanced tool for the case $g(0) \geq 0$ is given in appendix~\ref{App A}. From that analysis, point $E_6$ is stable whenever $g(0)>\frac{\xi}{4}dg(0)$, while it is a saddle for $g(0)=0$.
\end{itemize}

\begin{figure}
\centering
\subfigure[]{%
\includegraphics[width=6cm,height=4cm]{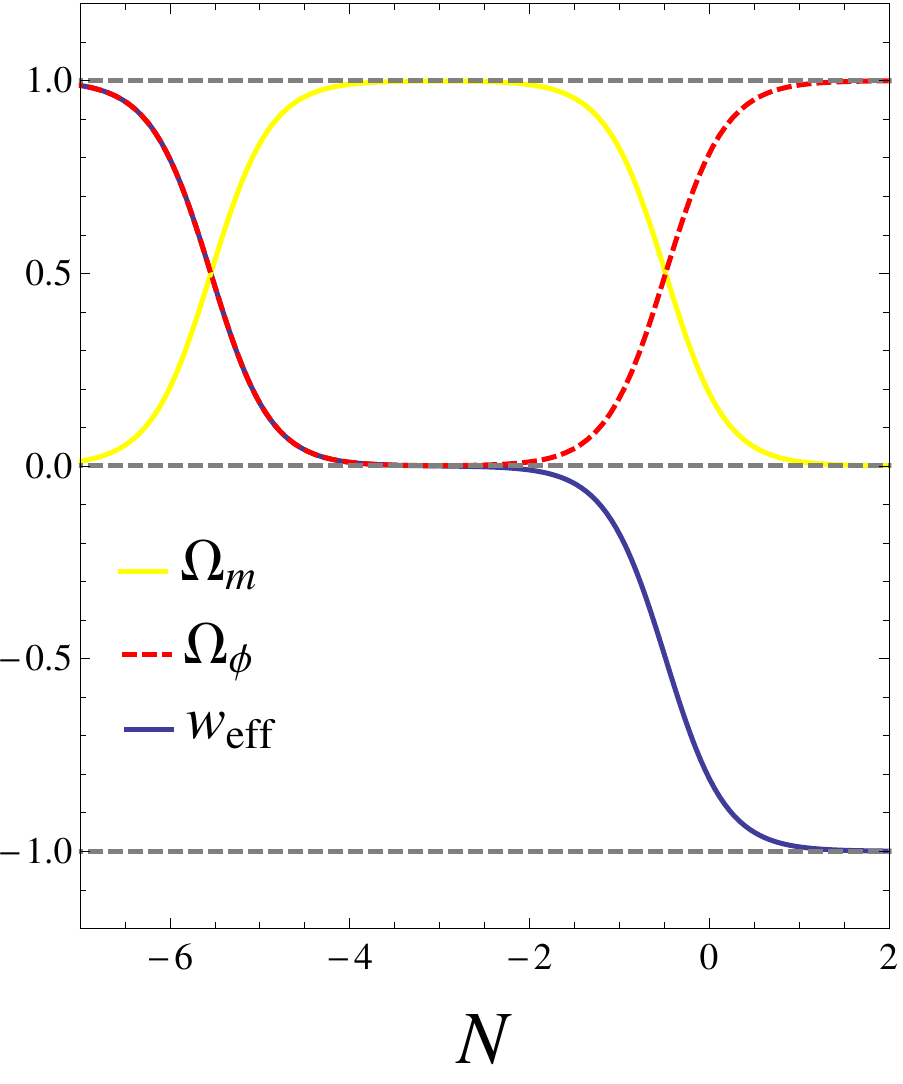}\label{weff_sinh_B}}
\qquad
\subfigure[]{%
\includegraphics[width=6cm,height=4cm]{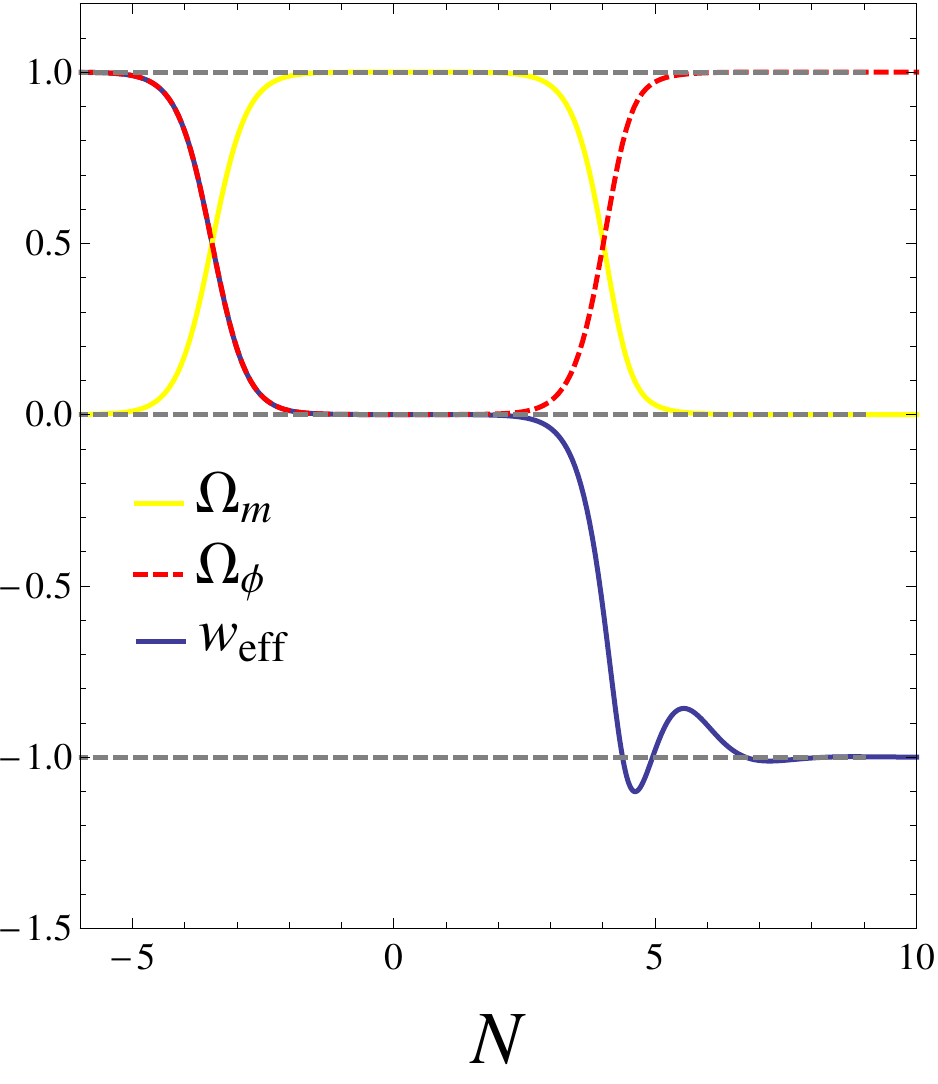}\label{weff_sinh_phantom_divide_II}}

\caption{Plot of $\Omega_m$, $\Omega_\phi$, $w_{\rm eff}$ versus $N$ of model II with potential $V=V_0\sinh^{-\eta}(\mu\phi)$. In both panels we have taken $w=0$, $\xi=10$, $\mu=1$, $\eta=2$, $\beta=0$ with different initial conditions.}\label{fig:weff_II_sinh}
\end{figure}

From the stability analysis of the critical points, depending on the choice of parameters and initial conditions, we observe that the universe starts from a stiff matter solution (points $E_{1\pm}$) and evolves either towards an accelerated, scalar field kinetic energy dominated late time attractors (points $E_{5\pm}$) or towards an accelerated, scalar field dominated late time attractor (point $E_6$), possibly passing through either a matter dominated solution (point $E_0$) or a scaling solution (point $E_2$).
In this model, it is thus possible that the universe at late times accelerates without being driven by the scalar field potential (point $E_{5\pm}$), yet with a cosmological constant behavior ($w_{\rm eff}=-1$).
In all cases, this model can successfully describe the observed matter dominated to DE dominated transition (see Fig.~\ref{weff_sinh_B} for an explicit example).
We also notice the possibility of crossing the phantom divide line, as explicitly shown in Fig.~\ref{weff_sinh_phantom_divide_II} for a specific scalar field potential.
This implies that in this model the scalar field characterizes a quintom scenario \cite{Cai:2009zp}, which cannot be obtained with uncoupled scalar (single) field models.
Finally note that point $E_2$ represents a matter scaling solution which can again provides deviations from $\Lambda$CDM at the perturbation level, although leaving the background dynamics unchanged (see Sec.~\ref{sec:perturbation}).

\subsubsection{$\beta=1$ case}
\label{subsec:beta1}

The points $E_0$, $E_{1\pm}$, $E_2$, $E_3$, $E_4$, $E_{5\pm}$ found in the $\beta=0$ case (given in table \ref{tab:der_B_0}) are critical points of the $\beta=1$ case as well, and their properties are unchanged.
For this reason they we will not be discussed again here.
The only difference is given by critical point $E_6$, which now becomes a non-isolated critical set $(0,1,z,0)$.
This set corresponds to an accelerated, scalar field dominated universe and will be denoted by $E_{6z}$.
It is a normally hyperbolic set when $g(0) \left(1-\frac{\xi z}{1-z}\right) \neq 0$. It is stable spiral if $4g(0)\left(1-\frac{\xi z}{1-z}\right)>1$, it is stable node if $0<4g(0)\left(1-\frac{\xi z}{1-z}\right)<1$, otherwise it is saddle.
Note that if $1-\frac{\xi z}{1-z}=0$, $E_{6z}$ reduces to point $E_4$.
The phenomenological aspect of the $\beta=1$ case are the same as the $\beta = 0$ case, except that the universe can reach a de Sitter final state in a finite amount of time, namely for a non vanishing value of $H$ (i.e.~for $z\neq 0$).

%%%%%%%%%%%%%%%%%%%%%%%%%%%%%%%%%%%%%%%%%%%%%%%%%%%%%%%%%%%%%%%%%%%%%%%%%%%%%%%%%%%%%%%%%%%%%%%%%%%%%%%%%%%%%%%%

\section{Cosmological perturbation and structure formation}\label{sec:perturbation}

\subsection{Linear perturbations and quasi-static approximation}

In this section, we focus on the behavior at the perturbation level of the interacting DE models considered in the previous sections.
We first present the general scalar perturbation equations at linear order, and then examine the effects arising in the process of cosmological structure formation within the quasi-static approximation.

At the background level we assume a spatially flat universe in agreement with cosmological observations, while at the linear perturbations level we work in the Newtonian gauge where the perturbed metric in Cartesian coordinates can be written as
\begin{equation}
{\rm d}s^2 = - (1+2\Phi) {\rm d}t^2+(1-2\Psi)\, a^2(t) \left({\rm d}x^2+{\rm d}y^2+{\rm d}z^2\right) \,.
\end{equation}
Since there are no anisotropic stresses in the considered coupled models \cite{Koivisto:2015qua}, we take into account the equality of the scalar perturbations $\Phi$ and $\Psi$ from the start.
In other words, in what follows we will replace everywhere $\Phi$ by $\Psi$ to simplify the equations.
In the matter sector, the physical quantities to be perturbed are the matter energy density $\rho$, the matter pressure $p$, the matter four-velocity $u_{\mu}$ and the scalar field $\phi$ as
\begin{equation}
\rho \mapsto \rho+\delta \rho \,, \qquad  p \mapsto p+\delta p \,, \qquad    u_\mu
\mapsto u_\mu + \delta u_\mu \,, \qquad  \phi \mapsto \phi+\delta \phi
\end{equation}
with
\begin{equation}
 \delta u_\mu = \left( -\Psi , \partial_i v \right) \,,
\end{equation}
where $v$ denotes the perturbed scalar velocity of the matter fluid.
Note that the symbols $\rho$, $p$, $u_\mu$ and $\phi$ denote background quantities.

The derivation of the complete linear perturbed cosmological equations of Scalar-Fluid theories have been first obtained in \cite{Koivisto:2015qua,Boehmer:2015ina}.
For the gradient (derivative) coupling scalar-fluid models considered in our analysis, the perturbed equations in Newtonian
gauge are given by %~\cite{Koivisto:2015qua}
\begin{equation}
    \left(-\frac{k^2}{a^2} - \rho - V \right) \Psi
    -3 H \dot\Psi  -\frac{1}{2} \delta\rho
    -\frac{1}{2} V' \delta \phi  -\frac{1}{2} \dot\phi \dot{\delta\phi}
=0 \,,
    \label{pert:1}
\end{equation}
\begin{equation}
     \dot\Psi  + H \Psi
     +\frac{1}{2} \left( \rho + p - n^2 \dot\phi \frac{\partial f}{\partial n}
\right) v
     -\frac{1}{2}  \dot\phi \delta \phi  = 0 \,,
     \label{pert:2}
\end{equation}
\begin{multline}
\!\!\!\!\!
    \ddot\Psi +4 H \dot\Psi  + \left[ 3 H^2 +2 \dot{H}
+\frac{1}{2}
\dot\phi^2 - \frac{1}{2} n^2 \dot\phi \frac{\partial f}{\partial n} \right] \Psi
    +\frac{1}{2} \dot\phi \left( 2 \frac{\partial f}{\partial n} + n
\frac{\partial^2
f}{\partial n^2} \right) \frac{n^2}{\rho+p} \delta\rho \\
\!\!\!\!\!\!\!\!\!\!\!\!\!\!\!\!\!\!
 - \frac{1}{2} \delta p
    +\frac{1}{2} \left( n^2 \dot\phi \frac{\partial^2 f}{\partial\phi \partial
n} + V'
\right) \delta \phi
    +\frac{1}{2} \left( n^2 \frac{\partial f}{\partial n} - \dot\phi \right)
\dot{\delta\phi}
=0 \,,
    \label{pert:3}
\end{multline}
\begin{multline}
    3 H \left(2 \frac{\partial f}{\partial n} + n \frac{\partial^2 f}{\partial n^2}
\right) \frac{n^2}{
\rho+p} \delta\rho
    + \left( 2 \ddot\phi + 6H \dot\phi - 3H n^2 \frac{\partial f}{\partial n} \right)
\Psi
    + \left( 4 \dot\phi -3 n^2 \frac{\partial f}{\partial n} \right) \dot\Psi  \\
    -\frac{k^2}{a^2} n^2 \frac{\partial f}{\partial n} v
    + \left( -\frac{k^2}{a^2} +3 H n^2 \frac{\partial^2 f}{\partial\phi\partial n} -
V'' \right) \delta \phi
    -3 H \dot{\delta\phi}  - \ddot{\delta\phi}    = 0 \,,
    \label{pert:4}
\end{multline}
\begin{equation}
    \frac{1}{\rho+p} \left( \dot{\delta\rho} +3H\delta p +3 H \delta \rho\right)
-\frac{k^2}{a^2}v  -
3 \dot\Psi  =0 \,,
    \label{pert:5}
\end{equation}
\begin{multline}
    \left( \frac{\partial\rho}{\partial n} -n \dot\phi \frac{\partial f}{\partial n}
\right) \dot{v}
    +n \left[ 3 H \dot\phi \left(\frac{\partial f}{\partial n} +n\frac{\partial^2
f}{\partial n^2}\right) -3 H \frac{\partial^2\rho}{\partial n^2} -\ddot\phi
\frac{\partial
f}{\partial n} -\dot\phi^2 \frac{\partial^2 f}{\partial\phi\partial n} \right] v
+\frac{1}{n} \delta p \\
    - \dot\phi \left( 2 \frac{\partial f}{\partial n} +n \frac{\partial^2 f}{\partial
n^2} \right) \frac{n}{\rho+p} \delta\rho +\frac{\partial\rho}{\partial n} \Psi
    - n \left( 3H \frac{\partial f}{\partial n} + \dot\phi \frac{\partial^2
f}{\partial\phi\partial n}
\right) \delta\phi
    -n \frac{\partial f}{\partial n} \dot{\delta\phi}  =0 \,.
    \label{pert:6}
\end{multline}

We will analyze the implications of the perturbation equations (\ref{pert:1})-(\ref{pert:6}) on structure formation when the quasi-static approximation is considered.
The scope is to investigate the evolution of matter overdensities $\delta$ (where $\delta$ is given by $\frac{\delta \rho}{\rho}$) in the comoving matter gauge.
In order to simplify the following equations, we introduce the background quantities \cite{Koivisto:2015qua}:
\begin{equation}\label{quan}
    X = n^2 \frac{\partial^2 f}{\partial\phi\partial n} \,, \qquad Y = n^2
\frac{\partial f}{\partial
n} \,, \qquad Z = n^3 \frac{\partial^2 f}{\partial n^2} \,.
\end{equation}
Moreover from now on we assume cold DM ($w=0$), with energy density proportionals to the fluid particle number density $n$ (namely $\rho \propto n$).
This implies that the sound speed square of the fluid in the rest frame of the field
\begin{equation} \label{cs2}
c_s^2 \equiv n\frac{\rho_{,nn}}{\rho_{,n}}\,,
\end{equation}
vanishes.
The evolution equation for matter overdensities in the comoving matter gauge is then given by~\cite{Koivisto:2015qua}
\begin{eqnarray} \label{dres}
\ddot{\delta} & + & \left[ 2H + \frac{ V' Y + Y'\left( 2Y-\dot{\phi}\right)
\dot{\phi}}{\rho + Y\left( Y- \dot{\phi}\right) }\right]\dot\delta
 =  \frac{\rho}{2} \left[ 1+\frac{Y\left( Y-\dot{\phi}\right)}{\rho}\right]^{-1}
\delta\,,
\end{eqnarray}
where a prime denotes differentiation with respect to $\phi$.

In what follows we are going to explore how Eq.~(\ref{dres}) behaves in the standard matter dominated and scaling solutions obtained for both models I and II in Sec.~\ref{sec:background}.
These solutions can in fact be used to consistently describe the structure formation era of the universe at the background level, but deviations from the standard $\Lambda$CDM dynamics might appear at the perturbations level.
Note that for all these solutions, being them critical points where $w_{\rm eff} = w = 0$, the Hubble parameter scales as
\begin{equation}
H=\frac{2}{3(1+w_{\rm eff})(t-t_0)} = \frac{2}{3(t-t_0)} \,,
\label{eq:H_matter}
\end{equation}
where $t_0$ is a constant of integration which we set to be $0$ for simplicity.
This is indeed the expected background evolution in a matter dominated universe.

\subsection{Model I}

The background quantities (\ref{quan}) can always be rewritten in terms of the dimensionless variables (\ref{variable1}).
Specifically for Model~I they can be expressed as
\begin{eqnarray}
X&=& -\gamma H \sigma^{1-2\alpha}y^{2\alpha} \Big(\alpha s^{\beta+1}+\beta s^{\beta-1} g(s)\Big) \,, \\
Y&=&-\xi H s^\beta y^{2\alpha}\sigma^{1-2\alpha} \,, \\
Z&=& 2 \xi H s^\beta y^{2\alpha} \sigma^{1-2\alpha} \,.
\end{eqnarray}
Note that now one can directly relate the growth rate evolution equation (\ref{dres}) with the coordinates of the critical points obtained from the background analysis in Sec.~\ref{sec:background}.

\subsubsection{$(0,0)$ model}

According to Sec.~\ref{subsec:00model}, for this model we have only one critical point corresponding to a matter scaling solution, namely point~$A_3$ (we will ignore the special case $s_*^2 = 3$ where also $A_2$ describes a matter scaling solution).
For this point, Eq.~(\ref{dres}) reduces to
\begin{eqnarray} %\label{dres_c0}
\ddot{\delta} & + & \frac{4}{3 t}\dot\delta
 =  {\frac {2}{{t}^{2} \left( 2{\xi}^{2}+3 \right) }} \delta.
\end{eqnarray}
This equation can also be written as
 \begin{eqnarray} \label{dres_c0N}
{\delta''} & + & \frac{1}{2}\delta'
 =   {\frac {9}{2\, \left( 2{\xi}^{2}+3 \right) }} \delta\,
\end{eqnarray}
where prime denotes derivative with respect to $N$, by using the relation $\frac{d}{d N}=\frac{1}{H}\frac{d}{d t}$.
The general solution of Eq.~(\ref{dres_c0N}) is given by
\begin{equation}\label{soln_pert}
\delta=C_1\,a^{m_+}+C_2\,a^{m_-},
\end{equation}
where $C_1$ and $C_2$ are two constant of integration, and
\begin{equation}
m_+= \frac{(-3 - 2  \xi^2 + \sqrt{225+156 \xi^2+4\xi^4})}{4 (3 + 2 \xi^2)},\,\quad  m_-= \frac{(-3 - 2  \xi^2 - \sqrt{225+156 \xi^2+4\xi^4})}{4 (3 + 2 \xi^2)}.
\end{equation}
We can notice that in this case the growth rate of matter perturbations depends only on the parameter $\xi$.
Moreover the quantity $m_-$ is always negative, and thus it does not contribute to the growth of perturbations.
For $m_+$ we have $0\leq m_+\leq 1$, which implies that the growth rate for this interacting DE model is smaller than that of the uncoupled case where $\delta \propto a$~\cite{Copeland:2006wr}, which is correctly recovered for $\xi = 0$.

\subsubsection{$(0,\frac{1}{2})$ model}

\begin{figure}
\centering
\subfigure[]{%
\includegraphics[width=6cm,height=3.5cm]{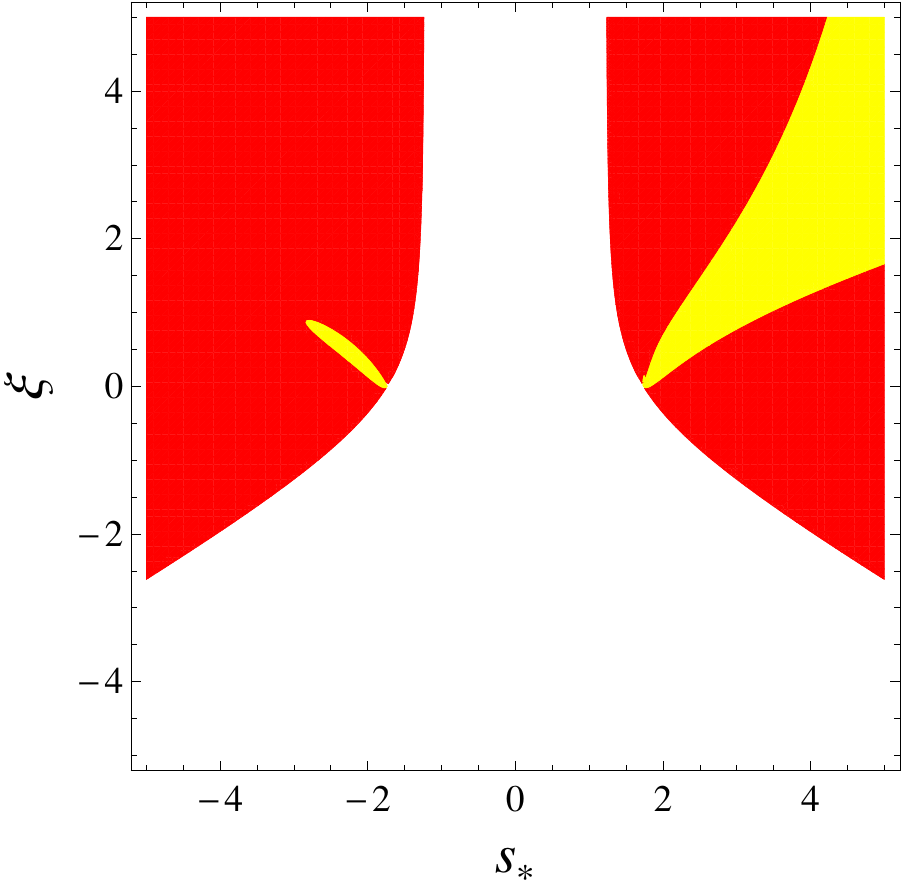}\label{fig:sca_pert_B8}}
\qquad
\subfigure[]{%
\includegraphics[width=6cm,height=3.5cm]{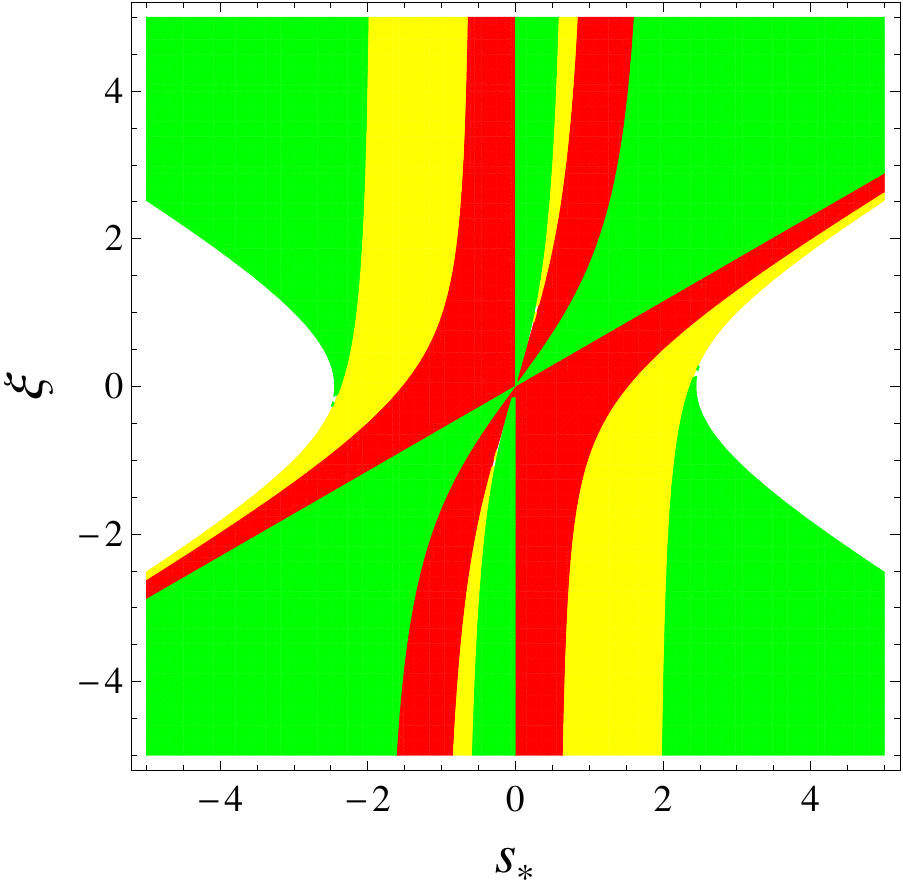}\label{fig:sca_pert_B9}}
\qquad
\subfigure[]{%
\includegraphics[width=6cm,height=3.5cm]{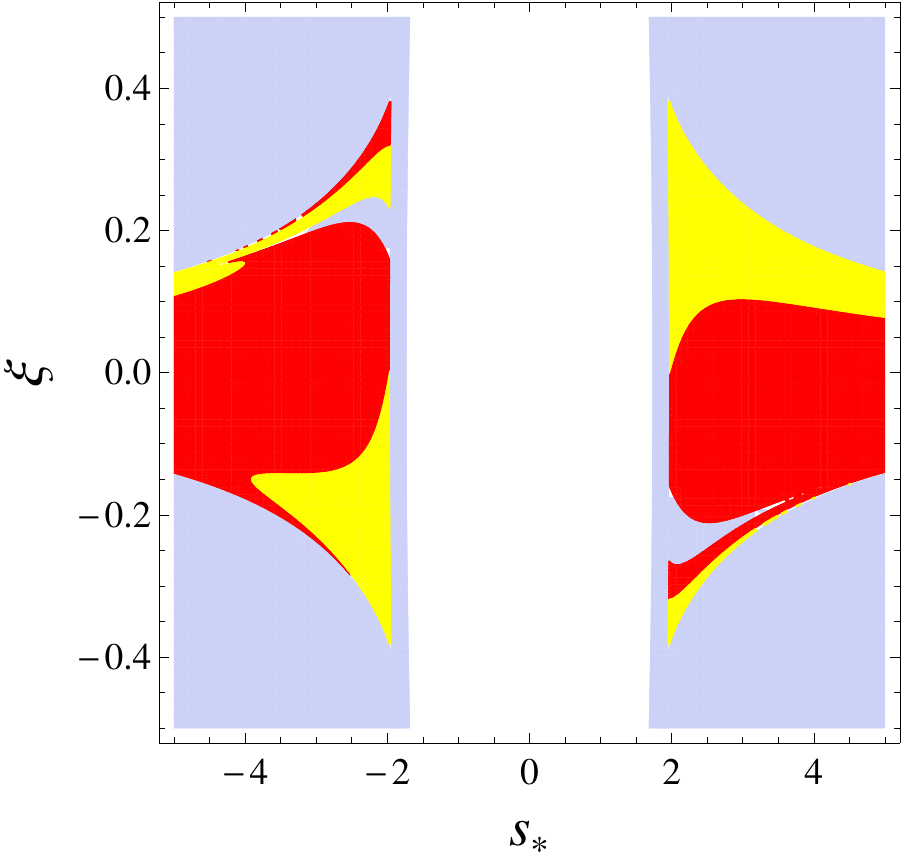}\label{fig:sca_pert_C3}}
\qquad
\subfigure[]{%
\includegraphics[width=6cm,height=3.5cm]{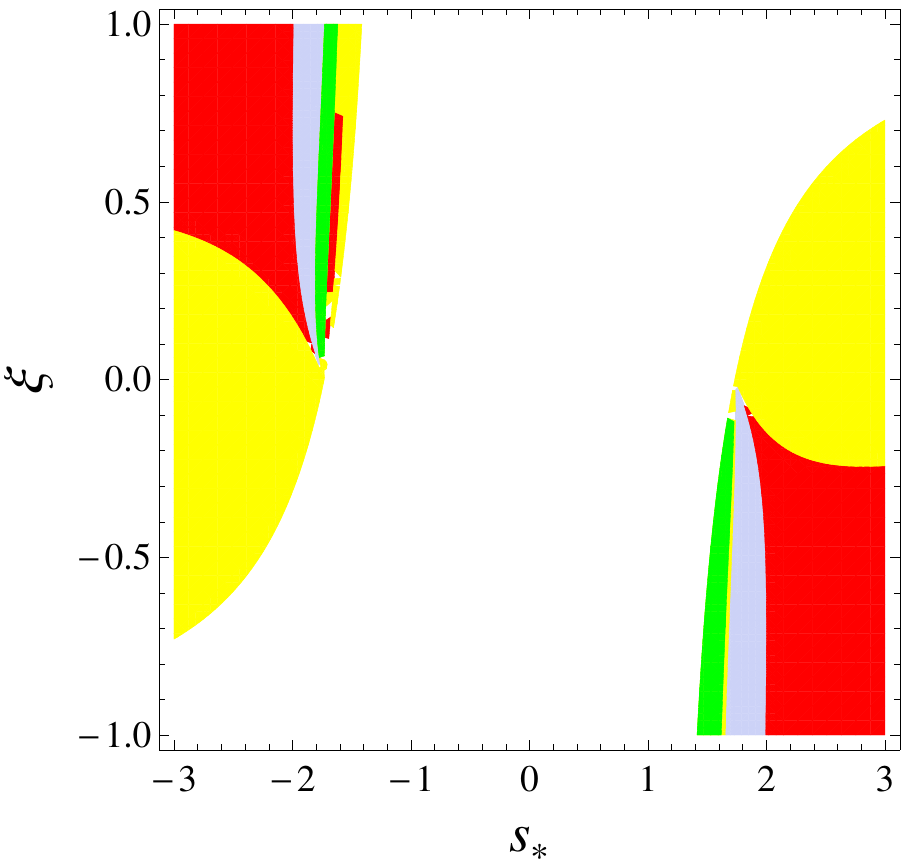}\label{fig:sca_pert_D2p}}
\qquad
\subfigure[]{%
\includegraphics[width=6cm,height=3.5cm]{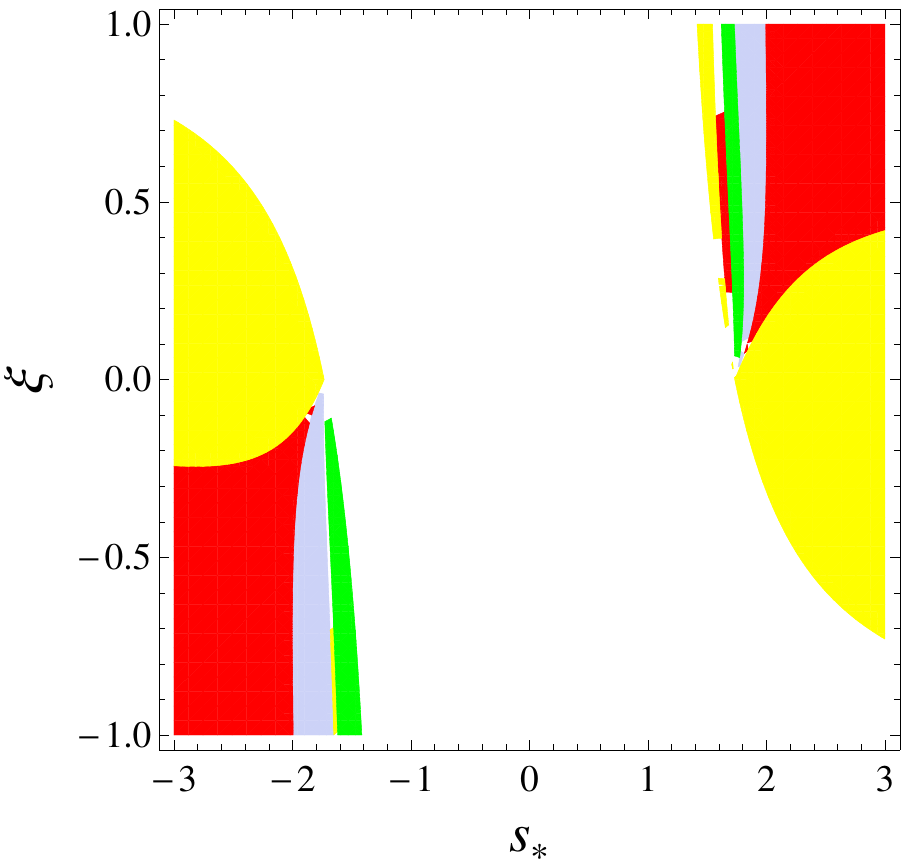}\label{fig:sca_pert_D2n}}
\qquad
\subfigure[]{%
\includegraphics[width=6cm,height=3.5cm]{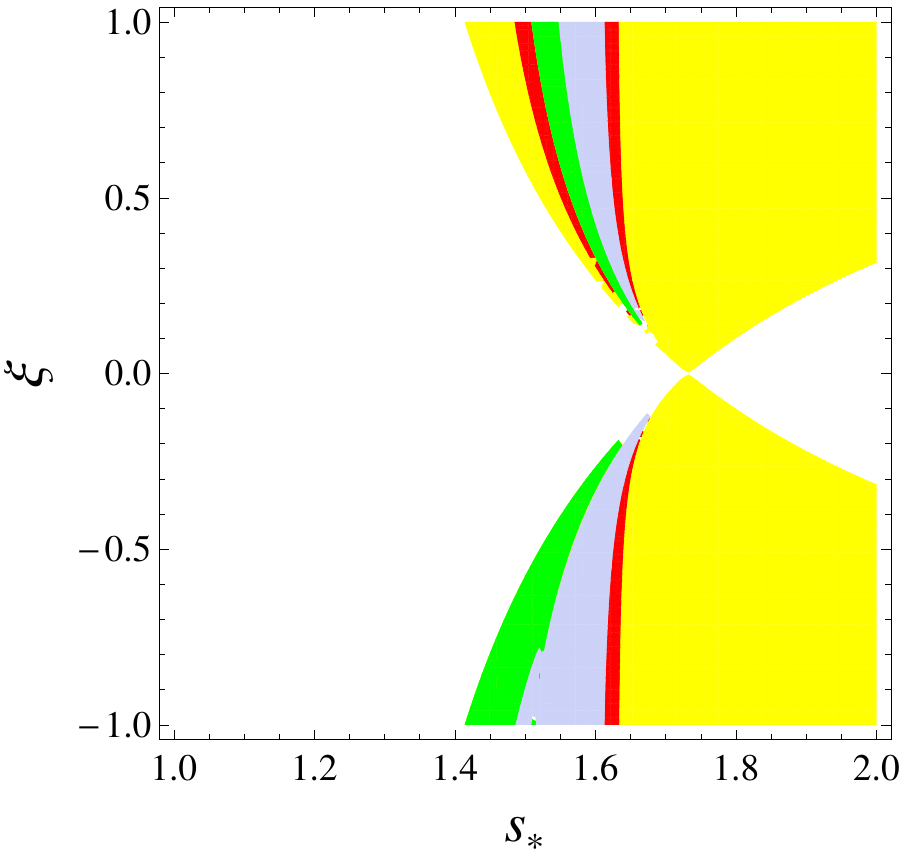}\label{fig:sca_pert_D4}}
\caption{The red regions in the $(s_*,\xi)$ parameter space denote the areas where $m_+>1$, the yellow regions denote the areas where $m_+<1$, the blue regions denote the areas where $m_+$ is imaginary (i.e.~matter perturbations are oscillating) and the green regions denote the areas where $m_+<0$.
Panels (a) and (b) correspond respectively to points $B_8$ and $B_9$ of the $(0,\frac{1}{2})$ model.
Panel (c) corresponds to point $C_3$ of the $(1,0)$ model.
Panels (d), (e) and (f) correspond respectively to points $D_{2+}$, $D_{2-}$ and $D_4$ of the $(1,\frac{1}{2})$ model.
}
\label{fig:region_sca_pert}
\end{figure}
 
This model coincides with the $k$-essence scalar field model studied in \cite{Nicola, Dutta:2016bbs} at the background level.
For this model, we obtain one standard matter dominated critical point $B_1$, one matter scaling solution $B_8$ and one scaling solution $B_9$ (see \cite{Dutta:2016bbs}).
For point $B_1$ the quantities $X$, $Y$, $Z$ vanish, hence Eq.~(\ref{dres}) reduces to that derived within the $\Lambda$CDM model, implying that there are no deviations from $\Lambda$CDM dynamics even at perturbation level.
For point $B_8$, the growth rate depends on the parameters $s_*$ and $\xi$, and the perturbation equation (\ref{dres}) becomes
\begin{eqnarray} \label{eq:pert_B8}
\ddot{\delta} & + &\frac{1}{3}\,{\frac { \left( \xi+\eta \right) \xi\, \left( 6\,s_*\,\xi-2\,
{\xi}^{2}+\sqrt {3} \right) -12(\,\eta\,\xi+2)+2(s_*-\xi) \left( 4\,
s_*+7\,\xi \right)}{t \left(  \left( {\xi}^{3}+\sqrt {3}\xi \right)  \left( \xi+\eta
 \right) -3(\,\eta\,\xi+2)+2({s_*}^{2}-{\xi}^{2} \right) }}
\dot\delta
 \nonumber \\&&~{} = \frac{1}{3}\,{\frac { \left( 3\,\eta\, \left( \xi+\eta \right) -2\,{s_*}^{
2} \right) ^{2}}{{s_*}^{2}{t}^{2} \left( 3\,\sqrt {3}\xi\, \left(
\xi+\eta \right)  \left( {\xi}^{2}+\sqrt {3} \right) -3(\eta\,\xi+2)+2
\,{s_*}^{2} \right) }} \delta,
\end{eqnarray}
where $\eta=\sqrt{\xi^2+2}$.
The solution of this equation has the same form as (\ref{soln_pert}), with $m_+$ and $m_-$ depending only on $s_*$ and $\xi$.
We will not report here the relation of $m_\pm$ with $s_*$ and $\xi$ due to their complicated and long expressions.
Nevertheless in Fig.~\ref{fig:sca_pert_B8} we have plotted the regions where the quantity $m_+>1$, i.e.~when the growth rate is enhanced by the coupling in comparison to that of uncoupled models, the regions where $m_+<1$, i.e.~when the growth rate is slower in comparison to that of uncoupled models.
We have checked that $m_+$ is not imaginary and that $m_-<1$ for any values of $s_*$ and $\xi$.
Finally for point $B_9$ we recover an equation similar to Eq.~\eqref{eq:pert_B8}, which however will not be shown due to the long expressions of its coefficients.
We can however mention that the quantity $m_+$ depends on both $s_*$ and $\xi$, while $m_-$ vanishes.
Also for this case, we have plotted in Fig.~\ref{fig:sca_pert_B9} the regions where the quantity $m_+>1$, the regions where $m_+<1$ and also the regions where $m_+<0$.

\subsubsection{$(1,0)$ model}

According to Sec.~\ref{subsec:10model}, for this model we have one standard matter dominated critical point $C_0$, one matter scaling solution $C_2$ and one scaling solution $C_3$.
For point $C_0$ the quantities $X$, $Y$, $Z$ vanish again, meaning that Eq.~(\ref{dres}) reduces to that obtained within the $\Lambda$CDM model.
For point $C_2$ Eq.~(\ref{dres}) reduces instead to 
\begin{eqnarray} %\label{dres_c0}
\ddot{\delta} & + & \frac{4}{3 t}\dot\delta
 =  \frac{2}{3}\,{\frac {2\,{s_*}^{2}{\xi}^{2}+1}{{t}^{2} \left( 2\,{s_*}^{2}{\xi}^{2}+3 \right) }} \delta\,
\end{eqnarray}
We obtain again the general solution as
\begin{equation}
\delta=C_1\,a^{m_+}+C_2\,a^{m_-},
\end{equation}
where now
\begin{align}
m_\pm = \frac{(-3 - 2 s_*^2 \xi^2 \pm \sqrt{81 + 204 s_*^2 \xi^2 + 100 s_*^4 \xi^4})}{4 (3 + 2 s_*^2 \xi^2)}\,.
\end{align}
Note that in this case the growth rate of matter perturbations depends only on the combination $s_*^2 \xi^2$.
Moreover we find again $m_- < 0$ and $0\leq m_+\leq 1$, which implies that the growth rate for this case is always smaller than that of uncoupled models.
Finally for point $C_3$ Eq.~(\ref{dres}) becomes
\begin{multline}
\ddot{\delta}  -\frac{2}{3 t}\,{\frac {{s_*}^{2} \left( -3\,\xi\,s_*\sqrt {-2\,\Omega} \left(
\Delta\,\sqrt {2}s_*-4 \right) +4\,{s_*}^{2}{\xi}^{2}+12 \right) }{
 \Theta \left( {s_*}^{2}{\xi}^{2}
\Omega-3\,\xi\,s_*\sqrt {-2\,\Omega}+3\,\Omega \right) }}
\dot\delta \\ =  -\frac{8}{3 t^2}\,\frac {s_*\Omega\,\sqrt {-2\,\Omega}}{ \Theta^{2} \left( \sqrt {-2\,\Omega}s_*{\xi}^{2}+{\frac {\sqrt {-2\,\Omega}}{s_*}}+6\,\xi \right)}
\delta\,,
\label{dres_c3}
\end{multline}
where 
\begin{align}
	\Theta &= \left( -\xi\,\sqrt {2\,{s_*}^{2}{\xi}^{2}+4\,{s_*}^{2}-2\,\Delta-12}s_*-{s_*}
^{2}{\xi}^{2}-2\,{s_*}^{2}+\Delta \right) \,,\\
	\Delta &=\sqrt{\xi^2 s_*^2\left((\xi^2+4)s_*^2-12\right)} \,,\\
	\Omega &=\Delta^2-2s_*^2+3 \,.
\end{align}
The growth rate results in a complicated expression depending on the parameters $s_*$ and $\xi$, however we have checked that $m_-<0$ for any $s_*$ and $\xi$.
In order to understand the behavior of the growth rate for some values of these parameters, in Fig.~\ref{fig:sca_pert_C3} we have numerically plotted the regions in the $(s_*,\xi)$ parameter space where the exponent $m_+$ is greater or smaller than 1.
This gives the regions in which matter overdensities grow faster or slower than in the uncoupled case.

\subsubsection{$(1,\frac{1}{2})$ model}

According to Sec.~\ref{subsec:112model}, for this model we obtain one standard matter dominated critical point $D_0$, two matter scaling solutions $D_{2\pm}$ and one general scaling solution $D_4$.
As expected for point $D_0$ the quantities $X$, $Y$, $Z$ vanish, implying that there are no deviations from standard $\Lambda$CDM dynamics even at the perturbations level. For point $D_{2+}$, Eq.~(\ref{dres}) becomes
\begin{multline} \label{dres_D2p}
\ddot{\delta}  - \frac{1}{3}\,{\frac { \left( \xi\,s_*+\Delta \right) s_*\,\xi\,
 \left( 2\,{s_*}^{2}{\xi}^{2}-6\,{s_*}^{2}\xi+9 \right) +6\,{
s_*}^{2} \left( \xi-4 \right)  \left( \xi+1 \right) +72}{t \left(
{s_*}^{3}{\xi}^{3} \left( \xi\,s_*+\Delta \right) +3\,{s_*
}^{2}{\xi}^{2}+6\,{s_*}^{2}-18 \right) }}
\dot\delta \\
 = \frac{1}{3}\,{\frac { \left( \Delta\,s_*\,\xi+{\Delta}^{2}-2\,{s_*}^{2
} \right) ^{2}}{{s_*}^{2}{t}^{2} \left( {s_*}^{4}{\xi}^{4}+
\Delta\,{s_*}^{3}{\xi}^{3}+2\,\Delta\,s_*\,\xi+5\,{\Delta}^{2}
+2\,{s_*}^{2}-36 \right) }}
 \delta\,,
\end{multline}
where $\Delta=\sqrt{\xi^2 s_*^2+6}$.
As in the previous cases, we have plotted in Fig.~\ref{fig:sca_pert_D2p} the regions on the $(s_*,\xi)$ parameter space where $m_+>1$, $m_+<1$, $m_+<0$ and where $m_+$ is imaginary.
We have also checked that either $m_-<1$ or imaginary.
A similar plot for point $D_{2-}$ is provided in Fig.~\ref{fig:sca_pert_D2n} (we do not present explicitly the equivalent to Eq.~\eqref{dres} for point $D_{2-}$), and again we either find $m_-<1$ or imaginary for every choice of the parameters.
Finally for point $D_4$ Eq.~\eqref{dres} becomes
\begin{multline} \label{dres_D4}
\ddot{\delta}-\frac {4 s_*^{4} \left( s_*^{2}\xi\, \left( 2\,s_*^{4}\xi-6\,s_*^
{4}-15\,s_*^{2}\xi+36\,s_*^{2}+63\,\xi-54 \right) -6\,\Theta \right) 
}{3 t \left( 2\,s_*^{6}{\xi}^{2}-6\,s_*^{4}{\xi}^{2}+3\,\Theta \right)}
\dot\delta \\
 = \frac {4 \, s_*^{4}{\Theta}^{2}}{ 3 {t}^{2}{\xi}^{2} \,\left( 2\,s_*^{4}+9\,s_*^{2}-18
 \right) ^{2} \left( 2\,s_*^{6}{\xi}^{2}-2\,s_*^{4}{\xi}^{2}+\Theta \right) }
 \delta\,,
\end{multline}
where $\Theta=(2\,{s_*}^{4}{\xi}^{2}-2\,s_*^{4}-3\,s_*^{2}{\xi}^{2}+12\,s_*^{2}-18)$.
For this case we have plotted as well in Fig.~\ref{fig:sca_pert_D4} the regions in the $(s_*,\xi)$ parameter space where $m_+>1$ (regions where matter over-densities grow faster than the uncoupled case), $m_+<1$ (regions where matter over-densities grow slower than the uncoupled case), $m_+<0$ and where $m_+$ is imaginary.
We have also checked numerically that $m_-$ is either $<1$ or imaginary for any values of $s_*$ and $\xi$.

\subsection{Model II}

\begin{figure}
\centering
\includegraphics[width=8cm,height=6cm]{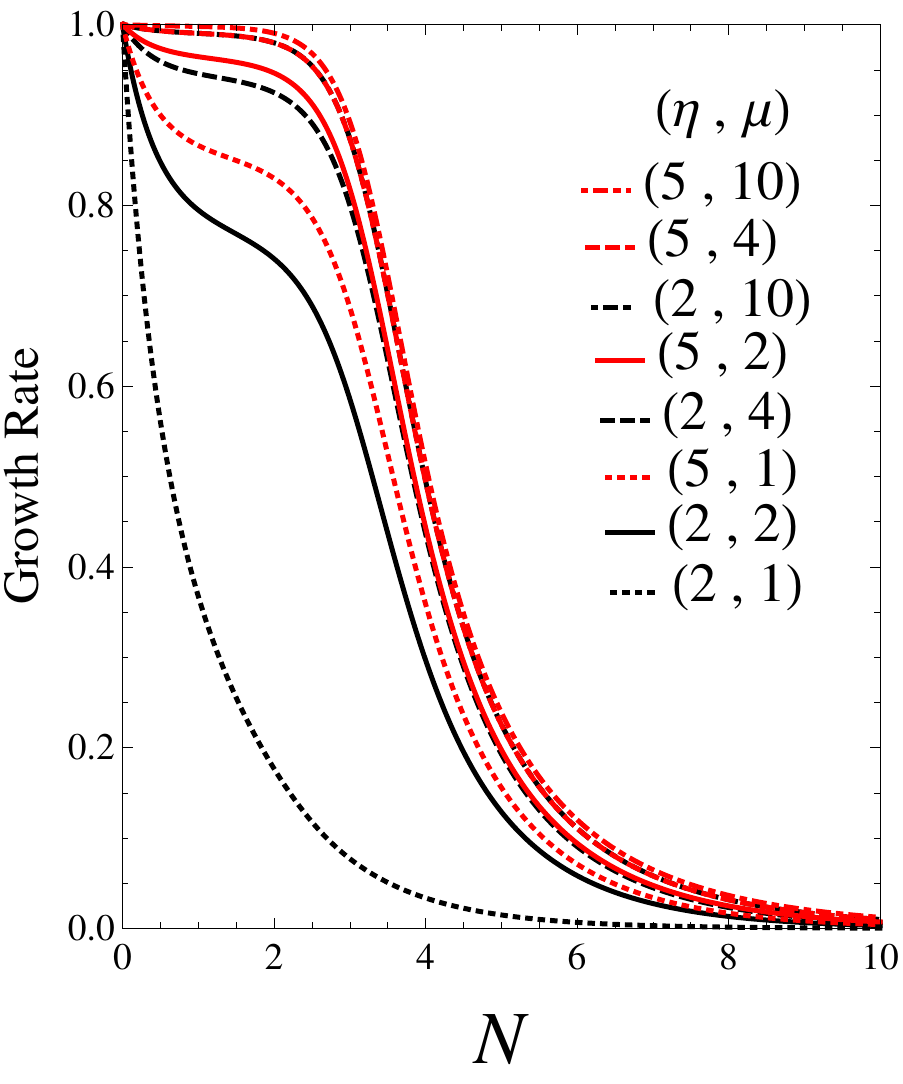}
\caption{Plot of the growth rate  $\frac{d \ln \delta}{d N}$ versus $N$ for Model~II with $\beta = 0$ and the scalar field potential $V=V_0\sinh^{-\eta}(\mu\phi)$ with $\xi=1$ and $V_0= H_0^2$.}
\label{fig:sca_pert_E2_sinh}
\end{figure}

From Eqs.~\eqref{quan} we find that for Model~II $X$ vanishes, while $Y=-\xi\, H_0$ and $Z=2\, \xi\, H_0$.
In both $\beta = 0$ and $\beta = 1$ cases, we obtain one standard matter dominated critical point $E_0$ and one matter scaling solution $E_2$ (cf.~Sec.~\ref{sec:model2}).
Since point $E_0$ is independent of the potential for its existence, its behavior at the perturbation level is the same as that derived in the case of an exponential potential.
This case has already been analysed in \cite{Koivisto:2015qua}, and thus will not been considered here.
Point $E_2$ instead depends on the choice of the scalar field potential, and thus its dynamics at the perturbation level will be different for different potentials.
In what follows we present an explicit example choosing the scalar field potential as
\begin{equation}
	V(\phi)  =V_0\sinh^{-\eta}(\mu\phi) \,,
\end{equation}
where $V_0$, $\eta$ and $\mu$ are parameters of suitable dimension.
For this potential we find $s_* = \pm \mu \eta$ and Eq.~(\ref{dres}) yields
\begin{multline}
\ddot{\delta} + \left[2H+\frac{3}{2}\frac {\eta\,\mu\,{\xi}^{2}H_0 \sqrt {1+\left( {\frac {9 H^2}{2 {s_*}^{2} V_0}} \right) ^{
\frac{2}{\eta}} }}{{H_0}^{2}{s_*}^{2}{\xi}^{2}+3\,{
H}^{2}{s_*}^{2}+3\,H H_0\,s_*\,\xi-9\,{H}^{2}}
\right] \dot{\delta} \\
= \frac{9}{2}\frac { \left( {s_*}^{2}-3 \right) ^{2}{H}^{4}}{{s_*}^{2
} \left( {H_0}^{2}{s_*}^{2}{\xi}^{2}+3\,{H}^{2}{s_*}^{2}+3\,H_0
s_*\,\xi-9\,{H}^{2} \right) } \delta \,.
\label{dres_E2}
\end{multline}
Note that the general solution of this equation is not of the form provided in Eq.~\eqref{soln_pert}, since the coefficients now are no longer constant being $H$ generally time dependent (cf.~Eq.~\eqref{eq:H_matter}).
We solve this equation numerically and present the evolution of the growth rate in Fig.~\ref{fig:sca_pert_E2_sinh}.
From this figure, we notice that at higher redshift (small $N$) and for sufficiently high values of $\eta$ and $\mu$, the matter overdensities grow at a constant rate, similar to the $\Lambda$CDM result where $d\ln\delta / dN = 1$ always.
For lower values of the parameters instead we find a growth rate which differs from the standard $\Lambda$CDM results even at early times.
Furthermore we have checked that for $\beta = 0$ if $\eta<0$ and $\mu>0$ the growth rate becomes $\gg 1$ and grows indefinitely as $N$ increases.
This reduces the region in $(\eta,\mu)$ parameter space which can give an evolution of matter perturbation comparable to the observed behavior of the universe.
A similar plot can be obtained for the $\beta=1$ choice, but in this case the growth rate is $\gg 1$ and grows indefinitely as $N$ increases when $\eta>0$ and $\mu<0$.

\section{Conclusion}\label{sec:conc}

In this paper we investigated the cosmological evolution of scalar field DE models with gradient coupling to the DM fluid, i.e.~where a coupling between the derivative of the scalar field $\partial_\mu\phi$ and the fluid's 4-velocity $u_\mu$ is present.
This coupling has been realised at the Lagrangian level by considering the consistent variational approach introduced in \cite{Boehmer:2015sha}.
The coupling term appearing in the Lagrangian has the general expression given by $\sqrt{-g}f(\mathfrak{n},\mathfrak{s},\phi)/\mathfrak{n}\, (u^\mu\partial_\mu \phi)$, where $f$ is an arbitrary function of all its arguments.
Generalising the analysis of \cite{Boehmer:2015sha}, we have considered two models corresponding to two distinct choices of the coupling function $f(\mathfrak{n},\mathfrak{s},\phi)$ (see Table~\ref{models}).

The first objective of the present paper was to investigate the background dynamics of these interacting DE models for arbitrary self interacting potentials.
For this analysis we have employed well-known dynamical systems methods, which allowed us to explore the complete cosmological evolution of the models considered.
In general we found that different interesting cosmological solutions can be obtained (see Sec.~\ref{sec:background}) from these models: the observed late-time transition from matter domination to DE domination, matter scaling solutions, late-time accelerating scaling solutions (useful to solve the cosmic coincidence problem) and even possible crossing of the phantom barrier.
This implies that these interacting models can produce interesting phenomenology at cosmological distances, able to reproduce the observed evolution of the universe, and possible deviations from it, at the background level.

Moreover for each of the models analysed we found matter scaling solutions able to successfully describe the matter dominated era at the background level, but possibly giving deviations in the dynamics of linear perturbations.
For this reason the second objective of our investigation was to analyse the cosmological dynamics at the perturbation level.
In particular we studied the formation of cosmological structures within the quasi-static approximation (see Sec.~\ref{sec:perturbation}).
All the scaling solutions found in the background analysis of Sec.~\ref{sec:background} present deviations from the standard $\Lambda$CDM dynamics during the growth of cosmological structure.
We have parametrized and discussed these deviations in Sec.~\ref{sec:perturbation}, and provided explicit examples to better understand the differences with the standard cosmological scenario.
These deviations can in principle be constrained by observational data, providing in turn constraints on the free parameters characterizing the interaction of DE with DM in these models.
The confrontation with observational data is however outside the scope of the present analysis and will be left as material for future work.

In conclusion we studied here the cosmological dynamics, at both background and perturbation levels, of interacting DE models where a gradient coupling between a scalar field and a matter fluid has been implemented using the variational approach introduced in \cite{Boehmer:2015sha}.
These DE models expand the possible theoretical arena where an interaction in the dark sector can be well defined at the fully covariant level, providing in this way consistent equations of motion at both background and perturbations cosmological levels.
The dynamics obtained from these equations can thus be effectively used to find deviations from the standard $\Lambda$CDM evolution which can eventually be tested against the observations, and thus used to constrain the parameter space of these interacting DE models.

\acknowledgments
J.D.~is thankful to IUCAA for warm hospitality and its facility for doing research work.
N.T.~acknowledges support from the Labex P2IO and an Enhanced Eurotalents Fellowship.

%%%%%%%%%%%%%%%%%%%%%%%%%%%%%%%%%%%%%%%%%%%%%%%%%%%%%%%%%%%%%%%%%%%%%%%%%%%%%%%%%%%%%%%%%%%%%%%%%%%%%%%%%%%%%%%%%%%%%%%%%%%%%%%%%%%%%%%%%%%%%%%%%%%%%%%%%%%%%%%%%%%%%%%%%%%%%%%%%%%%%%%%%%%%%%%%%%%%%%%%%%%%%%%%%%%%%%%%%%%%%%%%%%%%%%%%%%%%%%%%%%%%%%%%%%%%%%%%%%%%%%%%%%%%%%

\appendix

\section{Appendix: Center Manifold Theory (CMT)}

\subsection{General framework}

Without going into the mathematical background of center manifold theory, in the following we list some important steps on determining the dynamics of a center manifold near a critical point.
For more mathematical details and examples, we refer the reader to \cite{Wiggins,Perko}.

If the non-vanishing eigenvalues of the Jacobian matrix of a non-hyperbolic critical points have all negative real part (if at least one of them has positive real part then the point is unstable), then the stability of the critical point can be determined using CMT with the following operational steps:
\begin{itemize}
\item[1.] First translate the coordinates of the non-hyperbolic critical point under consideration to the origin and obtain a new set of autonomous equations in the new coordinates.
\item[2.] Express the non linear autonomous system of equations obtained in step 1 into the following standard form
           \begin{align}
               u'&=A u+f(u,v)\label{cmteqn1} \\
                v'&=B v+g(u,v)\label{cmteqn2}
            \end{align}
                    where $(u,v)\in \mathbb{R}^c \times \mathbb{R}^s$ with $f$ and $g$ satisfying
                     \begin{align*}
                          f(0,0)=0,\quad Df(0,0)=0\\
                          g(0,0)=0,\quad Dg(0,0)=0
                       \end{align*}
Here $A$ is a $c\times c$ matrix whose eigenvalues have zero real part, $B$ is $s\times s$ matrix whose eigenvalues have negative real part and $Df$ denotes the Jacobian matrix of $f$.
\item[3.] Determine a function $h(u)$, usually approximating it by a series expansion, which is at least $C^2$ and satisfies the following quasilinear partial differential equation
 \begin{align}\label{quasi}
 \mathcal{N}h(u)\equiv Dh(u)\left(A u+f(u,h(u))-B h(u)-g(u,h(u))\right)=0,
\end{align}
with $h(0)=0$ and $Dh(0)=0$.
\item[4.] The dynamics of the original system restricted to the center manifold is then determined by substituting the approximated solution of $h$ obtained in step 3 in the equation
 \begin{align}\label{cmtreduce}
 u'=A u+f(u,h(u))
 \end{align}
  The stability/instability of the system (\ref{cmtreduce}) implies the stability/instability of the original system.
  Note that usually Eq.~(\ref{cmtreduce}) reads $u'= k u^n$ for some constant $k$ and positive integer number $n$ (the lowest order in the expansion), for which stability is achieved only if $k < 0$ and $n$ is odd-parity, while any other case yield instability \cite{Perko}.
\end{itemize}
In what follows we explicitly show some examples of this analysis that have been used in the main body of the paper.
This should help the reader to better understand the practical application of the steps outlined above.

\subsection{Applications of CMT}
\label{App A}

\subsubsection*{Center manifold dynamics for point $C_{5}$ of model I with $\beta=1$, $\alpha=0$ when $g(0)=0$}

In this appendix we apply CMT to study the stability of point $C_{5}$ appearing in the analysis of Sec.~\ref{subsec:10model}. The coordinates of this point are $(x,y,s) = (0,1,0)$ (cf.~Table~\ref{tab:der_1_0}).
We first translate the point $(0,1,0)$ to the origin by using the transformation $x\rightarrow x$, $y\rightarrow y+1$, $s\rightarrow s$.
Then Eqs.~(\ref{x_der})-(\ref{s_der}) becomes
\begin{align}
 x' &= -\frac{1}{2} \Big(3 x ((w-1) x^2+(w+1)(y+1)^2+1-w) \nonumber\\
 & \qquad \qquad \qquad \qquad -\sqrt{6}(\xi\,\sqrt{1-x^2-(y+1)^2} s(x^2-1)+s (y+1)^2)\Big), \\
  y' &= -\frac{1}{2} (y+1) \left( 3 \left((w-1) x^2 +  (w+1) ((y+1)^2-1)\right)+\sqrt{6} x  (s-\xi\,\sqrt{1-x^2-(y+1)^2}s)\right),\\
  s'&=-\sqrt{6}\, x\, g(s).
\end{align}
Using the eigenvectors of the Jacobian matrix of the transformed system, we now introduce a new set of variables defined by
\[\left(\begin{array}{c}
X\\
Y\\
S \end{array} \right)=\left(\begin{array}{ccc}
1    & 0 &  -\frac{1}{\sqrt{6}}  \\
0    &1   & 0\\
0    & 0 &  1\\ \end{array} \right) \left(\begin{array}{c}
x\\
y\\
s \end{array} \right)\]
In terms of these new set of variables, the system of equations can now be written as
\[\left(\begin{array}{c}
X'\\
Y'\\
S' \end{array} \right)=\left(\begin{array}{ccc}
-3  & 0 & 0  \\
0  & -3 (w+1) & 0\\
0  & 0 & 0   \end{array} \right) \left(\begin{array}{c}
X\\
Y\\
S \end{array} \right)+\left(\begin{array}{c}
g_1\\
g_2\\
f \end{array} \right)\]
where $f,\,g_1,\,g_2$ are polynomials of degree greater than 2 in $(X,\,Y,\,S)$ with
\begin{eqnarray}
f(X,Y,S)&=&-\sqrt{6}g(S)\,X-g(S)\,S
\end{eqnarray}
whereas $g_1$, $g_2$ are not shown due to their lengths.
Now, the coordinates which correspond to non-zero eigenvalues $(X,Y)$ can be approximated in terms of $S$ by the expanded functions
\begin{equation}
h_1(S)=a_2 S^2+a_3 S^3+\mathcal{O}(S^4),
\end{equation}
\begin{equation}
h_2(S)=b_2 S^2+b_3 S^3+\mathcal{O}(S^4),
\end{equation}
respectively.
The quasilinear partial differential equation which the function vector
\[\mathbf{h}=\left(\begin{array}{c}
h_1\\
h_2\ \end{array} \right)\,,
\] 
has to satisfy, is given by
\begin{equation}\label{quasi_C}
D \mathbf{h(S)}\left[A S+\mathbf{F}(S,\mathbf{h}(S))\right]-B \mathbf{h}(S)-\mathbf{g}(S,\mathbf{h}(S))=\mathbf{0} \,,
\end{equation}
with \[\mathbf{g}=\left(\begin{array}{c}
g_1\\
g_2 \end{array} \right),~~~~~ \mathbf{F}=f, ~~~~~B= \left(\begin{array}{cc}
 -3 & 0 \\
 0 &-3(w+1) \end{array} \right),~~~~~ A=0. \]
In order to solve the Eq.~(\ref{quasi_C}), we substitute $A$, $\textbf{h}$, $\mathbf{F}$, $B$, $\mathbf{g}$ into it and equate equal powers of $S$ in order to obtain $\mathbf{h}(S)$ up to the desired order.
By comparing powers of $S$ from both sides of Eq.~(\ref{quasi_C}) we obtain the constants $a_2$, $a_3$, $b_2$, $b_3$ as
\begin{equation}
	a_2=-\frac{\sqrt{6}}{18}dg(0) \,,\quad
	a_3=\frac{\sqrt{6}}{36}\left(2\,dg(0)^2+d^2g(0)\right) \,,\quad
	b_2=-\frac{1}{12} \,,\quad
	b_3=-\frac{1}{18}\,dg(0) \,.
\end{equation}
Finally the dynamics of the reduced system is determined by the equation
\begin{equation}
S'=A\,S+\mathbf{F}(S,\mathbf{h}(S)),
\end{equation}
so that
\begin{align}%\label{CMT_S_prime_C+}
S'=-dg(0)S^2-\left(\frac{1}{3}\,dg(0)+\frac{1}{2}\, d^2g(0)\right)\,S^3
+\mathcal{O}(S^4).
\end{align}
Hence point $C_{5}$ is always unstable since at the lowest order we obtain a even-parity term.
If instead $dg(0)=0$ then the next term in the expansion must be considered, in which case the point is stable if $d^2g(0)>0$.

\subsubsection*{Center manifold dynamics for point $D_{5\pm}$ of model I with $\beta=1$, $\alpha=\frac{1}{2}$ when $g(0)=0$}

In this appendix we apply center manifold theory to study the stability of point $D_{5+}$ with coordinates $(x,y,s)=(0,1,0)$ and $D_{5-}$ with coordinates $(0,-1,0)$ for the cases where $g(0)=0$ (see Sec.~\ref{subsec:112model}).
We first analyze point $D_{5+}$.
We translate the point $(0,1,0)$ to the origin by using the transformation $x\rightarrow x$, $y\rightarrow y+1$, $s\rightarrow s$. Then Eqs.~(\ref{x_der})-(\ref{s_der}) becomes
\begin{align}
 x' &= -\frac{1}{2} \left(3 x ((w-1) x^2+(w+1)(y+1)^2+1-w)-\sqrt{6}(\xi\,(y+1)s(x^2-1)+s (y+1)^2)\right), \label{eq:A15} \\
  y' &= -\frac{1}{2} (y+1) \left( 3 \left((w-1) x^2 +  (w+1) ((y+1)^2-1)\right)+\sqrt{6} x  (s-\xi\,(y+1)s)\right),\\
  s'&=-\sqrt{6}\, x\, g(s) \,. \label{eq:A17} 
\end{align}
Using the eigenvectors of the stability matrix of the transformed system, we now introduce a new set of variables given by
\[\left(\begin{array}{c}
X\\
Y\\
S \end{array} \right)=\left(\begin{array}{ccc}
0                 & 1 &  0  \\
1    &0       &  \frac{(\xi-1)}{\sqrt{6}} \\
0    & 0 &  1\\ \end{array} \right) \left(\begin{array}{c}
x\\
y\\
s \end{array} \right)\]
In terms of these new set of variables, the system of equations \eqref{eq:A17}--\eqref{eq:A17} can now be written as
\[\left(\begin{array}{c}
X'\\
Y'\\
S' \end{array} \right)=\left(\begin{array}{ccc}
-3(w+1)  & 0 & 0  \\
0  & -3 & 0\\
0  & 0 & 0   \end{array} \right) \left(\begin{array}{c}
X\\
Y\\
S \end{array} \right)+\left(\begin{array}{c}
g_1\\
g_2\\
f \end{array} \right)\]
where $f,\,g_1,\,g_2$ are polynomials of degree greater than 2 in $(X,\,Y,\,S)$ with
\begin{eqnarray}
f(X,Y,Z,S)&=&-\sqrt{6}g(S)\,Y+g(S)\xi\,S-g(S)\,S
\end{eqnarray}
whereas $g_1$, $g_2$ are not shown due to their lengths.
Now the coordinates which correspond to non-zero eigenvalues $(X,Y)$ can be approximated in terms of $S$ by the functions
\begin{equation}
h_1(S)=a_2 S^2+a_3 S^3+\mathcal{O}(S^4),
\end{equation}
\begin{equation}
h_2(S)=b_2 S^2+b_3 S^3+\mathcal{O}(S^4),
\end{equation}
respectively. The quasilinear partial differential equation which the vector of functions
\[\mathbf{h}=\left(\begin{array}{c}
h_1\\
h_2 \end{array} \right)
\]
has to satisfy, is given by
\begin{equation}\label{quasi_D}
D \mathbf{h(S)}\left[A S+\mathbf{F}(S,\mathbf{h}(S))\right]-B \mathbf{h}(S)-\mathbf{g}(S,\mathbf{h}(S))=\mathbf{0} \,,
\end{equation}
where
\[\mathbf{g}=\left(\begin{array}{c}
g_1\\
g_2 \end{array} \right),~~~~~ \mathbf{F}=f, ~~~~~B= \left(\begin{array}{cc}
 -3(w+1) & 0 \\
 0 &-3 \end{array} \right), ~~~~~A=0. \]
In order to solve the Eq.~(\ref{quasi_D}), we substitute $A$, $\textbf{h}$, $\mathbf{F}$, $B$, $\mathbf{g}$ in it and equate equal powers of $S$ in order to obtain $\mathbf{h}(S)$ up the the desired order.
This process yields the values of the constants $a_2$, $a_3$, $b_2$, $b_3$ as
\begin{gather}
	a_2=-\frac{1}{12}(\xi-1)^2 \,, \qquad
	a_3=\frac{1}{18}(\xi-1)^3\,dg(0) \,, \qquad
	b_2=\frac{\sqrt{6}}{18}(\xi-1)^2\,dg(0) \,, \nonumber \\
	b_3=-\frac{\sqrt{6}}{72}(\xi-1)^2\left(4dg(0)^2(1-\xi)+2d^2g(0)+\xi\right) 
\end{gather}
Now the dynamics of the reduced system is determined by the equation
\begin{equation}
S'=A\,S+\mathbf{F}(S,\mathbf{h}(S)),
\end{equation}
which becomes
\begin{align}%\label{CMT_S_prime_C+}
S'=-dg(0)(1-\xi)S^2+\frac{1}{6}(\xi-1)\left(2dg(0)^2\,(1-\xi)+3 d^2g(0)\right)\,S^3
+\mathcal{O}(S^4).
\end{align}
This implies that point $D_{5+}$ is always unstable, unless either $dg(0)=0$ or $\xi = 1$.
If $dg(0)=0$ then the point is stable if $d^2g(0)(1-\xi)>0$, if instead $\xi = 1$ higher order terms in the expansion must be considered in order to determine stability.
A similar analysis can be performed for point $D_{5-}$, eventually leading to the equation
\begin{align}\label{CMT_S_prime_C-}
S'=-dg(0)(1+\xi)S^2-\frac{1}{6}(\xi+1)\left(2dg(0)^2\,(1+\xi)+3 d^2g(0)\right)\,S^3
+\mathcal{O}(S^4).
\end{align}
Consequently, the point $D_{5-}$ is unstable unless either $dg(0)=0$ or $\xi = - 1$, in which cases higher order terms must be analysed to determine its stability.

\subsubsection*{Center manifold dynamics for point $E_6$ of model II when $g(0)>0$}

In this appendix we apply center manifold theory to study the stability of point $E_6$ of Model~II with coordinates $(x,y,z,s)=(0,1,0,0)$ when $g(0)>0$ (see Sec.~\ref{subsec:beta0}).
First we translate the point $(0,1,0,0)$ to the origin by using the transformation $x\rightarrow x$, $y\rightarrow y+1$, $z\rightarrow z$, $s\rightarrow s$.
Then Eqs.~(\ref{x_der_B})-(\ref{s_der_B}) become
\begin{align}
x'&=-\frac{1}{2(z-1)}\Big[3 x (z-1)\left(1-w+(w-1)x^2+(1+w)(y+1)^2\right) \nonumber\\
& \qquad \qquad \qquad \qquad \qquad \qquad +\sqrt{6}\left(-\xi\,z(1-x^2)-s(z-1)(y+1)^2\right)\Big],\\
y'&=-\frac{(y+1)}{2(z-1)}\left[3(z-1)\left((1+w)((y+1)^2-1)+(w-1)x^2\right)+\sqrt{6}x\left((z-1)s+z\xi\right)\right],\\
z'&=\frac{z}{2}\left[3( z-1)\left((1+w)((y+1)^2-1)+(w-1)x^2\right)+\sqrt{6}z \xi x \right],\\
s'&=-\sqrt{6}\,x\,g(s).
\end{align}
Using the eigenvectors of the stability matrix of the transformed system, we now introduce a new set of variables given by
\[\left(\begin{array}{c}
X\\
Y\\
Z\\
S \end{array} \right)=\left(\begin{array}{cccc}
0                 & 1 &  0  &  0\\
-\frac{\sqrt{6}g(0)}{\sqrt{9+12g(0)}}          &0       &  \frac{1}{2}\frac{\xi \left(3+\sqrt{9+12g(0)}\right)}{\sqrt{9+12g(0)}} & \frac{1}{2}\frac{\left(3+\sqrt{9+12g(0)}\right)}{\sqrt{9+12g(0)}} \\
\frac{\sqrt{6}g(0)}{\sqrt{9+12g(0)}}                   & 0 &  \frac{1}{2}\frac{\xi \left(-3+\sqrt{9+12g(0)}\right)}{\sqrt{9+12g(0)}}                                            &  \frac{1}{2}\frac{\left(-3+\sqrt{9+12g(0)}\right)}{\sqrt{9+12g(0)}}\\
0                 & 0 & \xi                                                &  0 \end{array} \right) \left(\begin{array}{c}
x\\
y\\
z\\
s \end{array} \right)\]
In terms of these new set of variables, the system of equations can now be written as
\[\left(\begin{array}{c}
X'\\
Y'\\
Z'\\
S' \end{array} \right)=\left(\begin{array}{cccc}
-3(w+1)  & 0 & 0  &  0\\
0  & \frac{3}{2}\,{\frac {4\,g \left( 0 \right) -\sqrt {9+12\,g \left( 0 \right) }+
3}{\sqrt {9+12\,g \left( 0 \right) }}}
 & 0 & 0\\
0  & 0 &  -\frac{3}{2}\,{\frac {\sqrt {9+12\,g \left( 0 \right) }+4\,g \left( 0 \right)
+3}{\sqrt {9+12\,g \left( 0 \right) }}}
 &  0\\
0  & 0 & 0  & 0 \end{array} \right) \left(\begin{array}{c}
X\\
Y\\
Z\\
S \end{array} \right)+\left(\begin{array}{c}
g_1\\
g_2\\
g_3\\
f \end{array} \right)\]
where $f,\,g_1,\,g_2,\,g_3$ are polynomials of degree greater than 2 in $(X,\,Y,\,Z,\,S)$ which will not be written down due to their length.
At this point the coordinates which correspond to non-zero eigenvalues $(X,Y,Z)$ can be approximated in terms of $S$ by the expanded functions
\begin{equation}
h_1(S)=a_2 S^2+a_3 S^3+\mathcal{O}(S^4),
\end{equation}
\begin{equation}
h_2(S)=b_2 S^2+b_3 S^3+\mathcal{O}(S^4),
\end{equation}
\begin{equation}
h_3(S)=c_2 S^2+c_3 S^3+\mathcal{O}(S^4),
\end{equation}
respectively.
The vector composed by these functions, namely
\[\mathbf{h}=\left(\begin{array}{c}
h_1\\
h_2\\
h_3 \end{array} \right)\]
has to satisfy the following differential equation
\begin{equation}\label{quasi_E}
D \mathbf{h(S)}\left[A S+\mathbf{F}(S,\mathbf{h}(S))\right]-B \mathbf{h}(S)-\mathbf{g}(S,\mathbf{h}(S))=\mathbf{0}.
\end{equation}
with
\begin{gather}
\mathbf{g}=\left(\begin{array}{c}
g_1\\
g_2\\
g_3 \end{array} \right) \,, \qquad
\mathbf{F}=f \,, \qquad
A=0 \,, \nonumber \\
B= \left(\begin{array}{ccc}
 -3(w+1) & 0 & 0\\
 0 & \frac{3}{2}\,{\frac {4\,g \left( 0 \right) -\sqrt {9+12\,g \left( 0 \right) }+
3}{\sqrt {9+12\,g \left( 0 \right) }}} &  0\\
 0 & 0  &   -\frac{3}{2}\,{\frac {\sqrt {9+12\,g \left( 0 \right) }+4\,g \left( 0 \right)
+3}{\sqrt {9+12\,g \left( 0 \right) }}} \end{array} \right) \,. 
\end{gather}
Finally in order to solve Eq.~(\ref{quasi_E}), we substitute $A$, $\textbf{h}$, $\mathbf{F}$, $B$, $\mathbf{g}$ into it and equate equal powers of $S$ to obtain $\mathbf{h}(S)$ order by order.
On comparing powers of $S$ from both sides of Eq.~(\ref{quasi_E}) we obtain the constants $a_2$, $a_3$, $b_2$, $b_3$, $c_2$, $c_3$ as
$$
a_2=0 \,, \quad a_3=0 \,, \quad b_2={\frac {4\,g \left( 0 \right) +\sqrt {9+12\,g \left( 0 \right) }+3}{2\,\xi\, \left( 4\,g \left( 0 \right) +3 \right) }} \,, \quad c_2=-{\frac {\sqrt {9+12\,g \left( 0 \right) }-4\,g \left( 0 \right)
-3}{2\,\xi\, \left( 4\,g \left( 0 \right) +3 \right) }} \,,
$$
\begin{multline}
	b_3=-\frac{1}{2 g \left( 0 \right)  \left( 4\,g
 \left( 0 \right) +3 \right) ^{2}{\xi}^{2}} \Big[8\,dg  \left( 0 \right) \xi\,
 \left( g \left( 0 \right)  \right) ^{2}+4\, \left( g \left( 0
 \right)  \right) ^{2}\sqrt {9+12\,f \left( 0 \right) } \\
 +3\,dg \left( 0 \right) \xi\,\sqrt {9+12\,g \left( 0
 \right) }-16\, \left( g \left( 0 \right)  \right) ^{3}
 +18\,dg \left( 0 \right) \xi\,g \left( 0 \right) \\
 -9\,\sqrt
{9+12\,g \left( 0 \right) }g \left( 0 \right) -48\, \left( g \left( 0
 \right)  \right) ^{2}+9\,dg  \left( 0
 \right) \xi-27\,g \left( 0 \right)\Big] \,, \nonumber
\end{multline}
\begin{multline}
	c_3=\frac{1}{2\, g \left( 0 \right)  \left( 4\,g
 \left( 0 \right) +3 \right) ^{2}{\xi}^{2}}\Big[ \,-8\,dg  \left( 0 \right) \xi\,
 \left( g \left( 0 \right)  \right) ^{2}+4\, \left( g \left( 0
 \right)  \right) ^{2}\sqrt {9+12\,g \left( 0 \right) } \\
 +3\,dg \left( 0 \right) \xi\,\sqrt {9+12\,g \left( 0
 \right) } +16\, \left( g \left( 0 \right)  \right) ^{3} 
 -18\,dg  \left( 0 \right) \xi\,g \left( 0 \right) \\
 -9\,\sqrt{9+12\,g \left( 0 \right) }g \left( 0 \right) +48\, \left( g \left( 0
 \right)  \right) ^{2}-9\,dg \left( 0
 \right) \xi+27\,g \left( 0 \right)\Big] \,, \nonumber
\end{multline}
Finally the dynamics of the reduced system is determined by the equation
\begin{equation}
S'=A\,S+\mathbf{F}(S,\mathbf{h}(S)),
\end{equation}
which becomes
\begin{align}\label{CMT_S_prime_B}
S'=-\frac{3}{2}\,{\frac {4\,g
 \left( 0 \right)-\xi\,dg  \left( 0 \right) }{{\xi}^{2}g \left( 0 \right)  \left( 4\,g \left( 0
 \right) +3 \right) }}
 S^5 
+\mathcal{O}(S^6).
\end{align}
Hence for $g(0)>0$, point $E_6$ is stable when $g(0)>\frac{\xi}{4}\,dg(0)$.
Note how in this case the first non vanishing power in the $S$-expansion is $S^5$, suggesting a highly non-linear dynamics for the center manifold.

\subsubsection*{Center manifold dynamics for point $E_6$ of model II when $g(0)=0$}

\begin{figure}%[t!]
\centering
\subfigure[]{%
\includegraphics[width=6cm,height=4.5cm]{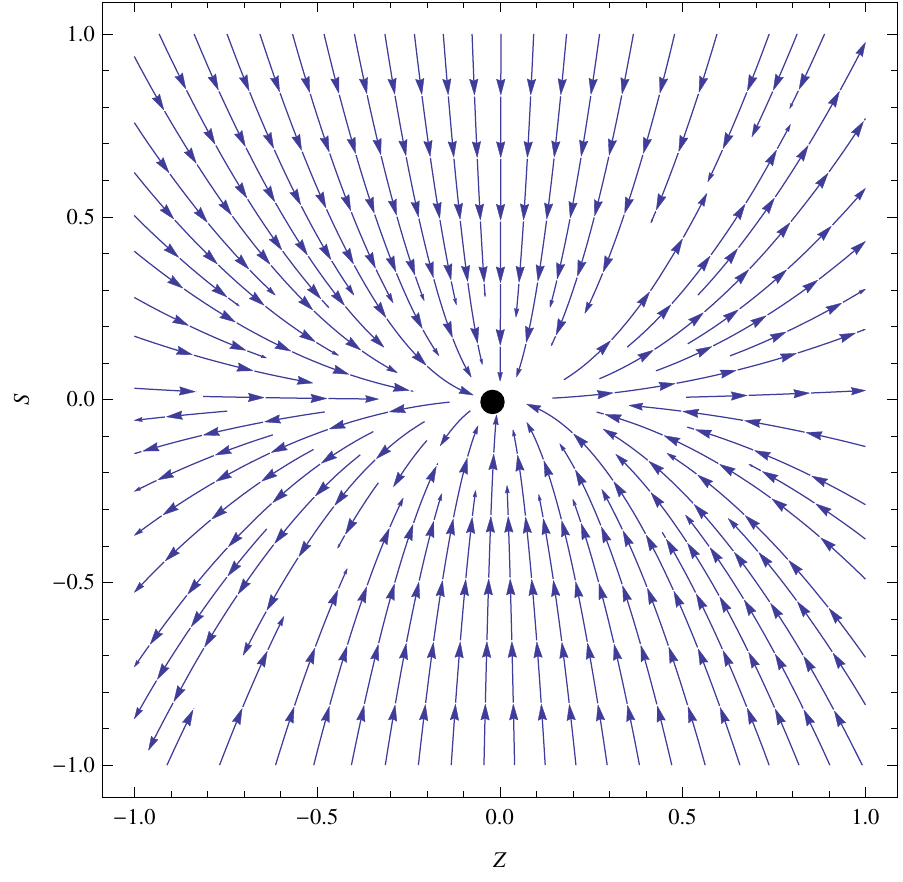}\label{streamplot_cmt_E6_pow_neg}}
\qquad
\subfigure[]{%
\includegraphics[width=6cm,height=4.5cm]{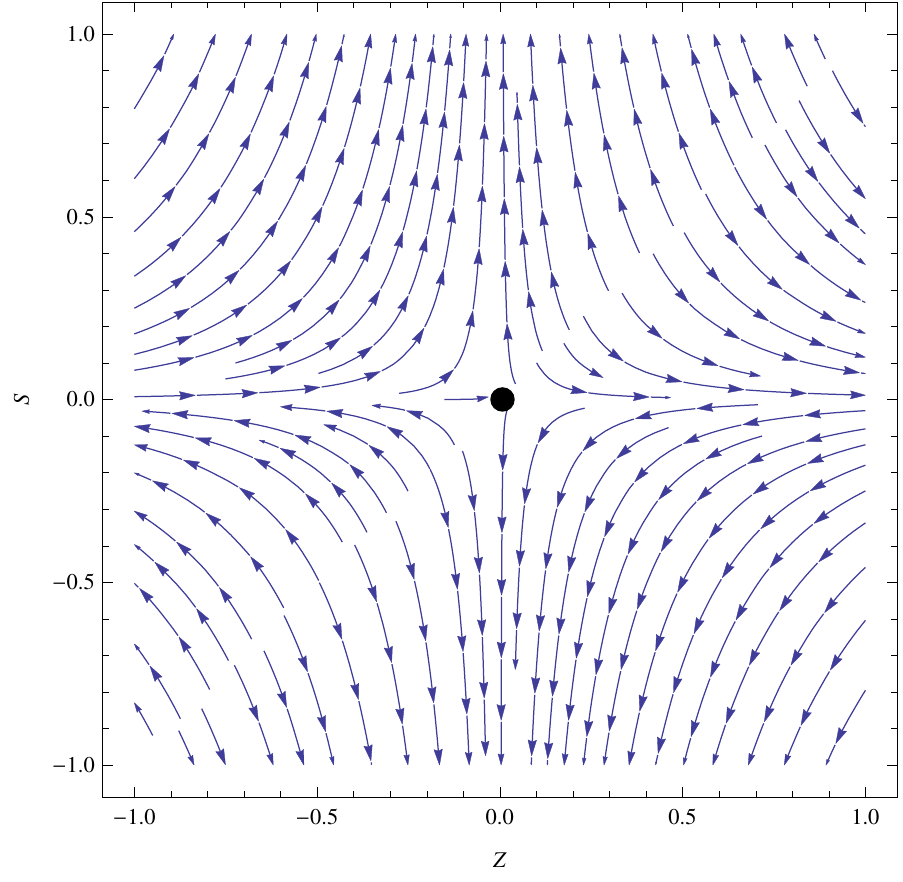}\label{streamplot_cmt_E6_pow_pos}}
\qquad
\subfigure[]{%
\includegraphics[width=6cm,height=4.5cm]{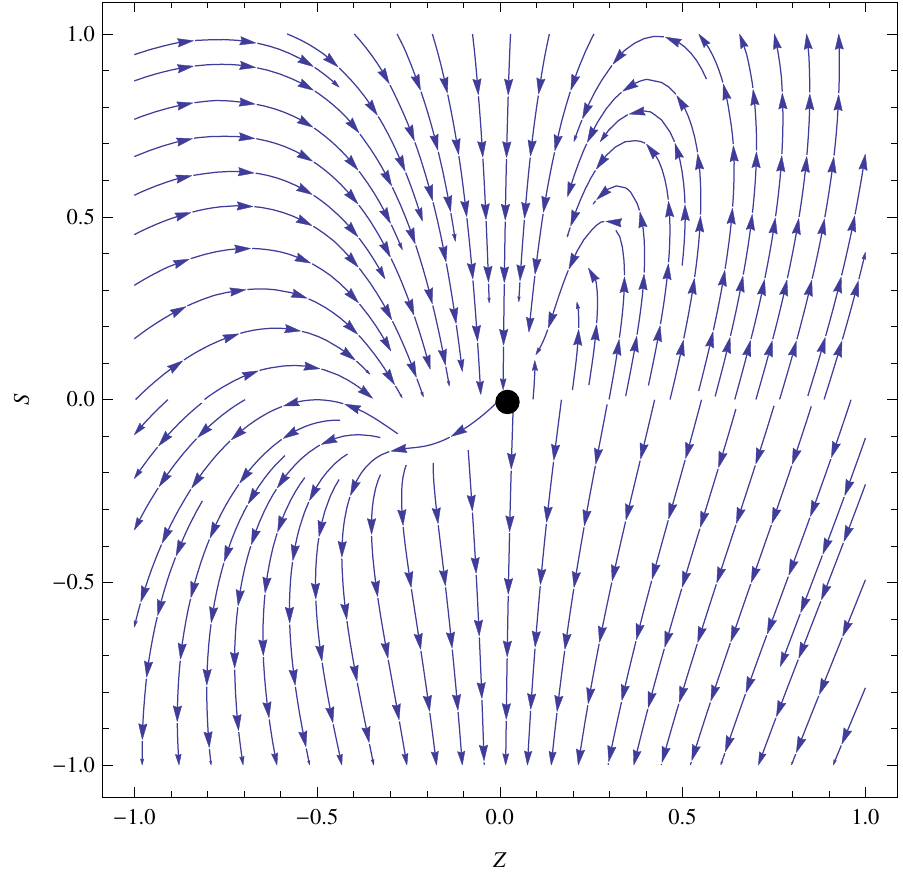}\label{streamplot_cmt_E6_nnp}}
\qquad
\subfigure[]{%
\includegraphics[width=6cm,height=4.5cm]{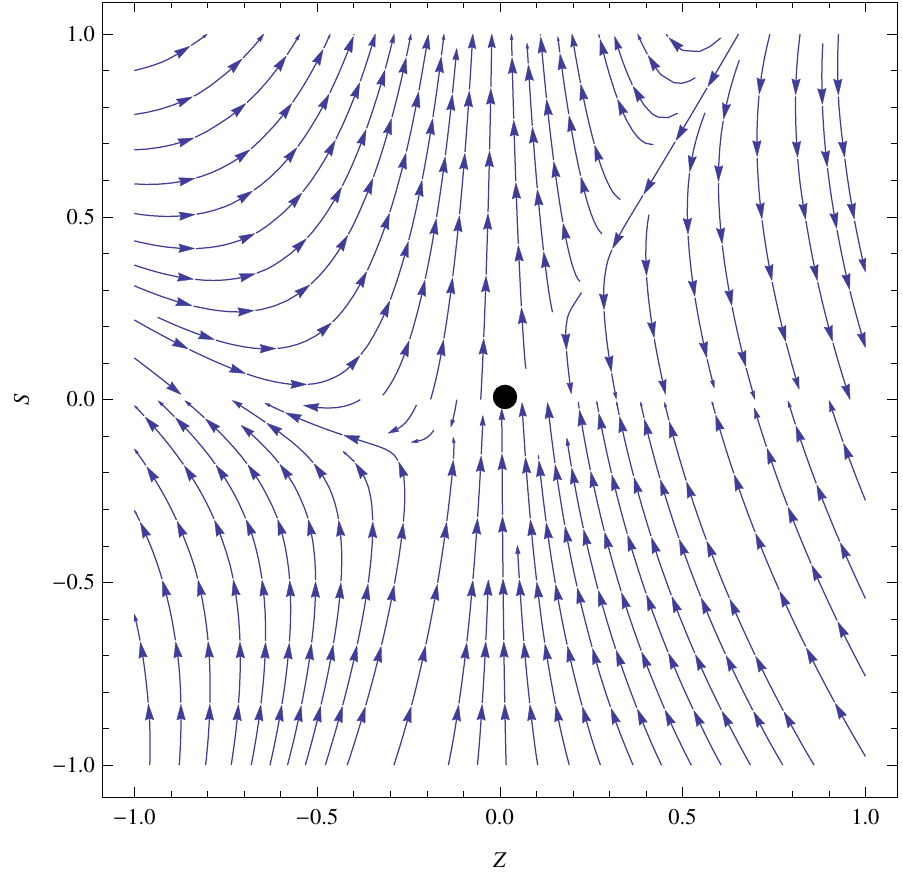}\label{streamplot_cmt_E6_npp}}
\caption{Phase portrait of the reduced system (\ref{CMT_E_Z_prime})-(\ref{CMT_E_S_prime}) in the $Z-S$ plane. In (a) we choose $\xi=-1$, $dg(0)=0$, $d^2g(0)=-2$, in (b)  $\xi=-1$, $dg(0)=0$, $d^2g(0)=2$, in (c) $\xi=-1$, $dg(0)=-1$, $d^2g(0)=1$, in (d) $\xi=-1$, $dg(0)=1$, $d^2g(0)=1$.}
\label{streamplot_cmt_E6}
\end{figure}

We finally explore the stability of point $E_6$ of model II (see Sec.~\ref{subsec:beta0}) in the case where $g(0)=0$.
The first step (translation of point $E_6$ to the origin) is the same as the previous example.
For the second step, using again the eigenvectors of the Jacobian matrix of the transformed system, we can introduce a new set of variables given by
\[\left(\begin{array}{c}
X\\
Y\\
Z\\
S \end{array} \right)=\left(\begin{array}{cccc}
1 & 0 &  \frac{\xi}{\sqrt{6}}  &  \frac{1}{\sqrt{6}}\\
0 &1  &  0 & 0 \\
0 & 0 &  1 &  0\\
0 & 0 & 0  &  1 \end{array} \right) \left(\begin{array}{c}
x\\
y\\
z\\
s \end{array} \right)\]
In terms of these new set of variables, the system of equations can now be written as
\[\left(\begin{array}{c}
X'\\
Y'\\
Z'\\
S' \end{array} \right)=\left(\begin{array}{cccc}
-3  & 0 & 0  &  0\\
0  &-3(1+w)& 0 & 0\\
0  & 0 & 0 &  0\\
0  & 0 & 0  & 0 \end{array} \right) \left(\begin{array}{c}
X\\
Y\\
Z\\
S \end{array} \right)+\left(\begin{array}{c}
g_1\\
g_2\\
f_1\\
f_2 \end{array} \right)\]
where $f_1,\,f_2,\,g_1,\,g_2$ are polynomials of degree greater than 2 in $(X,\,Y,\,Z,\,S)$ with
\begin{align}
f_1(X,Y,Z,S)&=-3(w+1)ZY+\frac{3}{2}(1-w)Z{X}^{2}-\frac{3}{2}(1+w)Z{Y}^{2
}+\frac{\sqrt {6}}{2} w \xi\,{Z}^{2}X \nonumber\\ & \quad
+3(1+w)\,{Z}^{2}Y -\frac{1}{4}(1+w){\xi}^{2}{Z}^{3} +\frac{\sqrt {6}}{2} (w-1) S\,Z\,X-\frac{1}{2}\,w\,\xi\,S\,{Z}^{2} \nonumber\\ & \quad
+\frac{1}{4}\,(1-w)\,{S}^{2}Z-\frac{3}{2}\,{Z}^{2}\,{X}^{2}(1-w)+\frac{3}{2}\,(w+1)\,{Z}^{2}\,{Y}^{2} +\frac{\sqrt
{6}}{2}\xi\,(1-w){Z}^{3}X \nonumber\\ & \quad
-\frac{1}{4}(1-w){\xi}^{2}{Z}^{4}+\frac{\sqrt {6}}{2}(1-w)S{Z}^{2}X -\frac{1}{2}(1-w)\xi\,S{Z}^{3} -\frac{1}{4}(1-w)\,{S}^{2}{Z}^{2} \,, \nonumber \\
f_2(X,Y,Z,S)&=-\sqrt {6}g \left( S \right) X+ g\left( S \right) \xi\,Z+g \left( S
 \right) S \,. \nonumber
\end{align}
whereas $g_1$, $g_2$ are not shown due to their lengths.
Now the coordinates over the 2D center manifold which correspond to the non-zero eigenvalues $(X,Y)$, can be approximated in terms of remaining coordinates $Z$, $S$ by the functions (up to third order)
\begin{align}
h_1(Z,S)&=a_1 Z^2+a_2 Z^3+a_3 Z S+a_4 Z^2 S+a_5 S^2+a_6 Z S^2+a_7 S^3 \,,\\
h_2(Z,S)&=b_1 Z^2+b_2 Z^3+b_3 Z S+b_4 Z^2 S+b_5 S^2+b_6 Z S^2+b_7 S^3 \,,
\end{align}
respectively.
Then the quasilinear partial differential equation which the function vector
\[\mathbf{h}=\left(\begin{array}{c}
h_1\\
h_2 \end{array} \right)\]
has to satisfy, is given by
\begin{equation}
D \mathbf{h(U)}\left[A U+\mathbf{F}(\mathbf{U},\mathbf{h}(\mathbf{U}))\right]-B \mathbf{h}(\mathbf{U})-\mathbf{g}(\mathbf{U},\mathbf{h}(\mathbf{U}))=\mathbf{0} \,,
\label{eq:A39}
\end{equation}
where
\[
\mathbf{U}=(Z,S), ~~~~\mathbf{g}=\left(\begin{array}{c}
g_1\\
g_2 \end{array} \right),~~~~~ \mathbf{F}=\left(\begin{array}{c}
f_1\\
f_2 \end{array} \right), ~~~~~B= \left(\begin{array}{cc}
 -3 & 0 \\
 0 & -3(1+w) \end{array} \right),~~~~~ A=0. 
\]
Equalling different powers of $S$ and $Z$ in Eq.~\eqref{eq:A39} yields finally the values of the constants $a_1$, $a_2$, $a_3$, $a_4$, $a_5$, $a_6$, $a_7$, $b_1$, $b_2$, $b_3$, $b_4$, $b_5$, $b_6$, $b_7$ as:
\begin{gather}
	a_1=-\frac{\xi}{\sqrt{6}} \,, \qquad a_2=-\frac{\xi}{\sqrt{6}} \,, \qquad a_3=\frac{\sqrt{6}}{18}\,\xi\, dg(0) \,, \nonumber \\
	a_4=\frac{\sqrt{6}}{18} \,\xi\, dg(0)(1-\frac{dg(0)\xi}{3})-\frac{\sqrt{6}}{36}\xi^2 \,, \qquad a_5=\frac{\sqrt{6}}{18} dg(0) \,, \nonumber \\
	a_6=-\frac{\sqrt{6}\xi}{108(1+w)}\left(8 dg(0)^2(w+1)-3d^2g(0)(w+1)+3(w-3)\right) \,, \nonumber \\
	a_7=-\frac{\sqrt{6}}{36(w+1)}\left(2dg(0)^2(w+1)-d^2g(0)(w+1)-4\right) \,,\nonumber\\
	b_1=-\frac{\xi^2}{12} \,, \qquad b_2=-\frac{\xi^2}{6} \,, \qquad b_3=\frac{1}{6}\,\frac{\xi(1-w)}{1+w} \,, \nonumber \\
	b_4=\frac{\xi}{18 (1+w)^2} \left(\xi\, dg(0)(w^2+2w-1)+3(1-w^2)\right), \qquad b_5=\frac{3-w}{12(w+1)} \,, \nonumber \\
	b_6=\frac{\xi\, dg(0)(w^2+w-3)}{9 (1+w)^2}, \qquad b_7=\frac{ dg(0)(w^2-5)}{18(1+w)^2}
\end{gather}
Finally the dynamics over the center manifold is determined by
\begin{eqnarray}
Z'&=&-\frac{\xi}{2}Z^2 S-\frac{1}{2}Z S^2,\label{CMT_E_Z_prime}\\
S'&=&dg(0)\,\xi\, ZS + dg(0) {S}^{2}+dg(0)\xi\, S\,{Z}^{2}+  \left( \frac{1}{2}\, d^2g(0)  - \frac{1}{3}\, dg(0)^2\, \right)\,\xi Z{S}^{2}\nonumber \\&&~{} +\left( \frac{1}{2}\, d^2g(0)-\frac{1}{3}\, dg(0)^2 \right){S}^{3}.\label{CMT_E_S_prime}
\end{eqnarray}
where $d^2g(0)$ denotes the second order derivatives of $g$ at $S=0$.
The full dynamics of the reduced 2D system (\ref{CMT_E_Z_prime})-(\ref{CMT_E_S_prime}) is complicated to determine analytically.
We can however check its stability for few examples.
In Fig.~\ref{streamplot_cmt_E6} we have plotted the phase portrait on the $Z-S$ plane choosing some values for the parameters $\xi$, $dg(0)$ and $d^2g(0)$.
As one can see from the figure, the point $E_6$ is always saddle for these choices of parameters.
Moreover, we have checked that for other choices of $\xi$, $dg(0)$ and $d^2g(0)$ point $E_6$ is still saddle.
This suggests that point $E_6$ is a saddle for any combination of the parameters.

\end{document}